\documentclass[fleqn,usenatbib]{mnras}

\usepackage{newtxtext,newtxmath}
\usepackage[T1]{fontenc}
\usepackage{amsmath}
\usepackage[utf8]{inputenc}
\usepackage{graphicx}
\usepackage{float}
\usepackage[version=4]{mhchem}
\usepackage{multirow}
\usepackage{verbatim}
\usepackage{orcidlink}

\title[Ground-based Characterization of L 98-59 d]{Ground-based Atmospheric Characterization of Super-Earth L 98-59 d at High Spectral Resolution}

\author[C.J. Cheverall et al.]{
Connor J. Cheverall\,\orcidlink{0009-0005-4330-7197},$^{1,2,3}$\thanks{E-mail: connor.cheverall@mail.udp.cl}
Nikku Madhusudhan\,\orcidlink{0000-0002-4869-000X},$^{3}$
Savvas Constantinou\,\orcidlink{0000-0001-6839-4569},$^{3}$
and P. R. McCullough$^{4,5}$
\\
$^{1}$Instituto de Estudios Astrofísicos, Facultad de Ingeniería y Ciencias, Universidad Diego Portales, Av. Ejercito Libertador 441, 8320000, Santiago, Chile\\
$^{2}$Centro de Excelencia en Astrofísica y Tecnologías Afines (CATA), Camino El Observatorio 1515, Las Condes, Santiago, Chile\\
$^{3}$Institute of Astronomy, University of Cambridge, Madingley Road, Cambridge CB3 0HA, UK\\
$^{4}$The William H. Miller III Department of Physics and Astronomy, Johns Hopkins University, Baltimore, MD 21218, USA\\
$^{5}$Space Telescope Science Institute, 3700 San Martin Dr., Baltimore, MD 21218, USA
}

\date{Accepted 2026 March 01. Received 2026 March 01; in original form 2025 August 20}
\pubyear{\the\year{}}
\begin{document}
\label{firstpage}
\pagerange{\pageref{firstpage}--\pageref{lastpage}}
\maketitle

\begin{abstract}

Atmospheric characterization of exoplanets using ground-based high-resolution transmission spectroscopy has traditionally focussed on large and close-in planets, such as hot Jupiters. In this work, we aim to extend this technique to smaller and more temperate planets by studying the atmospheric composition of the temperate super-Earth planet L 98-59 d ($\sim$$1.5\,\mathrm{R_{\oplus}}$; $\sim$$1.9\,\mathrm{M_{\oplus}}$). Using high-resolution transmission spectra obtained using IGRINS on the Gemini-South telescope, we demonstrate the feasibility for atmospheric characterization of super-Earths using ground-based facilities, and confirm the previous tentative JWST inference of hydrogen sulfide (\ce{H2S}) in the atmosphere of L 98-59 d at $\lesssim$3.9$\,\sigma$ ($B\sim390$). This is the first ground-based inference of a molecular species in the atmosphere of a super-Earth planet, and reveals the sensitivity of spectrographs on 8\,m-class telescopes to the atmospheric characterization of such planets. By exploring a grid of atmospheric models, we find that the data favors a cloud-free atmosphere with an abundance of \ce{H2S} corresponding to $\sim$1--10\,$\times$ solar metallicity. We additionally place constraints on the atmospheric abundances of other molecular species. Assuming cloud-free models, super-solar abundances for \ce{CH4} and \ce{NH3} are ruled out at 3.6$\sigma$ and $4.6\sigma$, respectively. Our results are consistent with previous suggestions that L 98-59 d is a super-Earth with possible disequilibrium production of \ce{H2S} driven by volcanic outgassing from the surface. Future studies combining multiple observations with different facilities may be able to further constrain the atmospheric composition of this planet. This work underscores the promise of  atmospheric characterization of super-Earth exoplanets using high-resolution spectroscopy with ground-based facilities.

\end{abstract} 


\section{Introduction} \label{intro} 

Driven by the unprecedented capabilities of JWST, exoplanetary science has entered an exciting new age of studying lower-mass exoplanets, with atmospheric characterization emerging as the frontier of this revolution. Low-mass planets have become a focal point of current JWST studies, attracting significant attention due to their potential for groundbreaking discoveries, e.g. the possibility of detecting biosignatures, and the uncertainty surrounding their internal structures \citep[e.g][]{madhusudan_carbon, benneke2024, holmberg2024}. In addition to sub-Neptunes, transmission spectra have also been observed for several super-Earth planets with JWST \citep[e.g.][]{may2023, lustig-yaeger2023, lim2023, moran2023, kirk2024, damiano, gressier2024, banerjee2024, bello-arufe2025}. However, placing strong constraints on the atmospheric compositions of such planets has faced various challenges, including small planetary radii and atmospheric scale heights, the potential presence of clouds/hazes, and high stellar activity from M-dwarf host stars.

In this regard, the natural complement to lower-resolution space-based spectroscopy is ground-based spectroscopy at high resolution \citep[e.g.][]{snellen_orbital_2010, brogi_signature_2012, birkby_detection_2013, brogi_rotation_2016, hoeijmakers_atomic_2018, gibson2020, giacobbe_five_2021}. The ability of this technique to resolve both transmission and emission spectra into individual spectral lines can break the degeneracy associated with broad and overlapping molecular bands, allowing for chemical identifications with increased confidence, and enable constraints on atmospheric dynamics \citep[e.g.][]{snellen_orbital_2010, brogi_rotation_2016, flowers2019, alonso-floriano_multiple_2019, beltz2021}. Ground-based observations at high resolution also offer unparalleled opportunities to probe the atmospheres of temperate, cloudy planets \citep{gandhi_seeing_2020, hood_prospects_2020} possibly in the habitable zones of their host stars. They may additionally be better placed to break degeneracies associated with the high stellar activity typical for M-dwarf host stars \citep{genest_effect_2022}. Studying low-mass exoplanets at high resolution may therefore enable us to gain novel insights into their atmospheric compositions and atmospheric dynamics, and make indirect inferences regarding their internal structures and evolutionary histories. 

As a result, high-resolution studies of low-mass planets have started to gain momentum in recent years. The atmospheres of a number of sub-Neptune planets have been observed \citep[e.g.][]{crossfield2011, deibert2019, cabot2024, dash2024, parker2025}, including attempts to characterize the atmospheres of super-Earth planets \citep{esteves2017, jindal2020, deibert2021, ridden-harper2023}. Despite this, we are aware of no molecular detections as yet in the atmospheres of super-Earth planets using ground-based facilities.
For the more temperate of these planets on longer-period orbits, ground-based studies are additionally and particularly complicated by the required telluric correction, more so than for their closer-in counterparts. In the near-infrared (NIR), which contains numerous spectral features of the molecules expected to be common in \ce{H2}-rich atmospheres, telluric correction has typically relied on a significant change in the Doppler shifts of the planetary spectral lines throughout the transit, although recent works have aimed to address this \citep[e.g.][]{cheverall2024}.

In this work, we study the atmosphere of the temperate super-Earth planet L 98-59 d using high-resolution transmission spectroscopy in the NIR. L 98-59 d (TOI-175 d) was discovered using the Transiting Exoplanet Survey Satellite mission \citep[TESS;][]{Ricker2015, kostov2019} and has since been followed-up with ground-based radial velocity observations \citep[e.g.][]{cloutier2019, demangeon2021}. This planet is one of three transiting planets orbiting a bright M-dwarf star ($K=7.10$\,mag, $M_{\mathrm{star}}=0.273\,\mathrm{M_{\odot}}$, $R_{\mathrm{star}}=0.303\,\mathrm{R_{\odot}}$, $T_{\mathrm{eff}}=3415$\,K), in a system which has previously been identified as promising for atmospheric characterization \citep{pidhorodetska2021}. L 98-59 d has a mass of $1.94\pm0.28$ $\mathrm{M_{\oplus}}$ and a radius of $1.52^{+0.12}_{-0.10}$ $\mathrm{R_{\oplus}}$ \citep[][see system parameters in Table \ref{system_parameters}]{demangeon2021}. The bulk parameters and equilibrium temperature \citep[$T_{\mathrm{eq}}=416\pm20$\,K; ][]{demangeon2021} for this planet lead to a degeneracy in the likely internal composition and structure, placing the planet intermediate between a pure rocky/silicate composition, and a gas rich one retaining the primary H-He envelope accreted from the disk \citep{rogers2015, j-rogers2021, luque-palle-2022}. A significant volatile layer is required to account for the bulk density of the planet \citep[$2.95^{+0.79}_{-0.51}$\,$\mathrm{gcm^{-3}}$;][]{demangeon2021}, introducing degeneracy between competing internal structure solutions, such as a rocky planet with a thick \ce{H2}-rich atmosphere, or a water world with a thinner atmosphere. In the latter case, possible water world scenarios may include an equal composition of rock and ice \citep[e.g.][]{luque-palle-2022} or a Hycean planet \citep[with a thin \ce{H2}-dominated atmosphere above the deep ocean; e.g.][]{madhusudhan2021}, although it is possible that strong stellar fluxes in the extreme-UV and X-ray spectral regions may have driven water loss from the planets in this system \citep{fromont2024}.
This ambiguity is one motivator for spectroscopic studies of exoplanet atmospheres, which can provide additional constraints to help inform the internal structure modelling of sub-Neptune and super-Earth planets and contribute toward breaking the degeneracy between possible internal structure solutions.

Spectroscopic studies of L 98-59 d have been conducted previously using space-based instrumentation. First, \cite{zhou2023} used the Hubble Space Telescope (HST) to observe this planet in transmission and found results inconsistent with a cloud-free, low-mean-molecular-weight atmosphere dominated by \ce{H2}. More recently, atmospheric characterization studies have been conducted for this planet using JWST NIRSpec G395H \citep{gressier2024, banerjee2024}. These works obtain a transmission spectrum which deviates from a flat line by 2.6--5.6\,$\sigma$ depending on the data reduction \citep{gressier2024}, and attribute this to a metal-rich planetary atmosphere. In addition, retrieval analysis tentatively identified the presence of prominent sulfur species hydrogen sulfide \ce{H2S} and sulfur dioxide \ce{SO2}. If confirmed, these sulfur species could be indicative of active volcanism  \citep[e.g.][]{claringbold2023, tsai2024}, although photochemistry could also provide a production mechanism for these species \citep{hu2013}. Such volcanic outgassing may be driven by the non-zero eccentricity of the planet \citep{demangeon2021}, which may lead to strong tidal heating \citep{seligman2024}. Recent interior-atmosphere modelling may support the role of tidal heating in prolonging magma oceans for the close-in planets in this system \citep{nicholls2025}, with an \ce{SO2}-rich volcanic atmosphere having recently been inferred for L 98-59 b \citep{bello-arufe2025}. Although previously thought unlikely that sulfur species such as \ce{H2S} could survive in detectable quantities in the atmospheres of terrestrial planets due to photochemical destruction \citep{hu2013}, it has since been suggested that such species may not be photochemically depleted above the cloud layer for planets orbiting M-dwarf stars, such as L 98-59 d \citep{jordan2021}. These results and possible active volcanism highlight L 98-59 d as an appealing target for further observation.

In the present work, we study the atmospheric chemical composition of L 98-59 d using high-resolution transmission spectroscopy.
We search for spectral signatures of various molecules expected to be prominent in temperate \ce{H2}-dominated exoplanetary atmospheres,
and confirm the presence of \ce{H2S}. 
To our knowledge, this is the first inference of a molecule in the atmosphere of a super-Earth using ground-based facilities, as well as the first ground-based inference of a sulfur-bearing species in the atmosphere of any planet. In addition, given the small change in the radial velocity of the planet during the transit complicating the telluric correction, the success of these observations builds upon \cite{cheverall2024}, where a proof-of-concept study for high-resolution transmission spectroscopy of temperate planets was conducted. We outline our methods in Section \ref{methods} before presenting our initial results in Section \ref{results}. In Section \ref{diversity} we aim to place constraints on the atmospheric composition of L 98-59 d. We summarize our work in Section \ref{discussion}.

\section{General Methods} \label{methods}

We here outline the analysis methods used in this work to identify molecular spectral signatures in the exoplanet atmosphere using high-resolution transmission spectroscopy. Each step is described in this section in turn, including the initial cleaning and normalization of spectra, the telluric correction, and finally the cross-correlation or likelihood-based model comparison in order to constrain the atmospheric composition and parameters.

\subsection{Observations} \label{observations}

We analyze time-series observations of the super-Earth L 98-59 d obtained using the IGRINS high-resolution spectrograph \citep{yuk_preliminary_2010, park_design_2014} as part of GO Program GS-2021A-Q-216 (PI: P.R. McCullough). IGRINS studies the NIR spectral range at a resolution of $R \sim$45,000. The wavelength coverage spans $\sim$1.45 -- 2.45 $\mu$m, split into 53 spectral orders, where several molecules expected to be common in \ce{H2}-dominated atmospheres have strong spectral features. At the time of these observations, IGRINS was situated on the 8\,m Gemini-South telescope (Cerro Pachón, Chile). 

Observational details for this program can be found in Table \ref{observing-info}. We primarily analyze observations taken on the night of 12th March 2021. Observations from a second night (10th February 2021) were also analyzed, however the number of available frames was limited due to the increased exposure times used, and the observations were at increased airmass. As a result, despite the increased median S/N per exposure due to longer exposure times, we find that the spectra observed on this night have limited sensitivity to the chemical species considered in this work, and so we do not include them in the main analysis (cross-correlation results and sensitivity testing for this night of observation can be found in Appendix \ref{appendix-night2}).

\begingroup
\setlength{\tabcolsep}{6pt}
\renewcommand{\arraystretch}{1.4}
\begin{table}
\centering
\begin{tabular}{lc}
\hline \hline
 Parameter & Value\\ \hline
 \hline
 \textbf{Planetary Parameters} \\
 $P$ & $7.4507245^{+0.0000081}_{-0.0000046}$ d\\
 $T_{0}$ & $2460121.1125^{+0.000119}_{-0.0001205}$ BJD$_{\mathrm{TDB}}$\\
 $M_{\mathrm{p}}$ & $1.94\pm0.28$ $\mathrm{M_{\oplus}}$ \\
 $R_{\mathrm{p}}$ & $1.521^{+0.119}_{-0.098}$ $\mathrm{R_{\oplus}}$ \\
 $a$ & $0.0486^{+0.0018}_{-0.0019}$ au \\
 $i$ & $88.449^{+0.058}_{-0.111}$ $^{\circ}$ \\
 $e$ & $0.074^{+0.057}_{-0.046}$ \\
 $b$ & $0.922\pm0.059$ \\
 $\omega$ & $180^{+27}_{-50}$ \\
 $T_{14}$ & $0.84^{+0.15}_{-0.20}$ h \\
 $T_{23}$ & $0.51^{+0.23}_{-0.18}$ h \\
 $K_{\mathrm{p}}$ & $70$ $\mathrm{km\,s^{-1}}$ \\
 \hline
 \textbf{Stellar Parameters} \\
 $M_{\mathrm{star}}$ & $0.273\pm0.030$ $\mathrm{M_{\odot}}$ \\
 $R_{\mathrm{star}}$ & $0.303^{+0.026}_{-0.023}$ $\mathrm{R_{\odot}}$ \\
 $T_{\mathrm{eff}}$ & $3415\pm135$ K \\
 $q_{1}$ & 0.09 \\
 $q_{2}$ & 0.214 \\
 $V_{\mathrm{sys}}$ & $-5.57851^{+0.00072}_{-0.00069}$ $\mathrm{km\,s^{-1}}$ \\
 \hline
\end{tabular} 
\caption{Planetary and stellar parameters used in this work for L 98-59 d. Values are adopted from \protect\cite{demangeon2021}, except the mid-transit time $T_{0}$ which is adopted from \protect\cite{gressier2024} and the quadratic limb-darkening coefficients $q_{1}$ and $q_{2}$ which are calculated using the open source Python package \texttt{exoctk} \citep[Exoplanet Characterization Toolkit;][]{bourque_2021}. Additionally, the radial velocity semi-amplitude $K_{\mathrm{p}}$ is derived in this work from 
the other system parameters: $K_{\mathrm{p}} = K_{\mathrm{star}}M_{\mathrm{star}}/M_{\mathrm{p}}$, where $K_{\mathrm{s}} = (2\pi G/P) ^{1/3} \times M_{\mathrm{p}}\sin(i)(M_{\mathrm{p}}+M_{\mathrm{s}})^{-2/3} (1-e^2)^{-1/2}$.
}
\label{system_parameters}
\end{table}
\endgroup

\begin{figure*}
    \centering
    \includegraphics[width=0.49\linewidth]{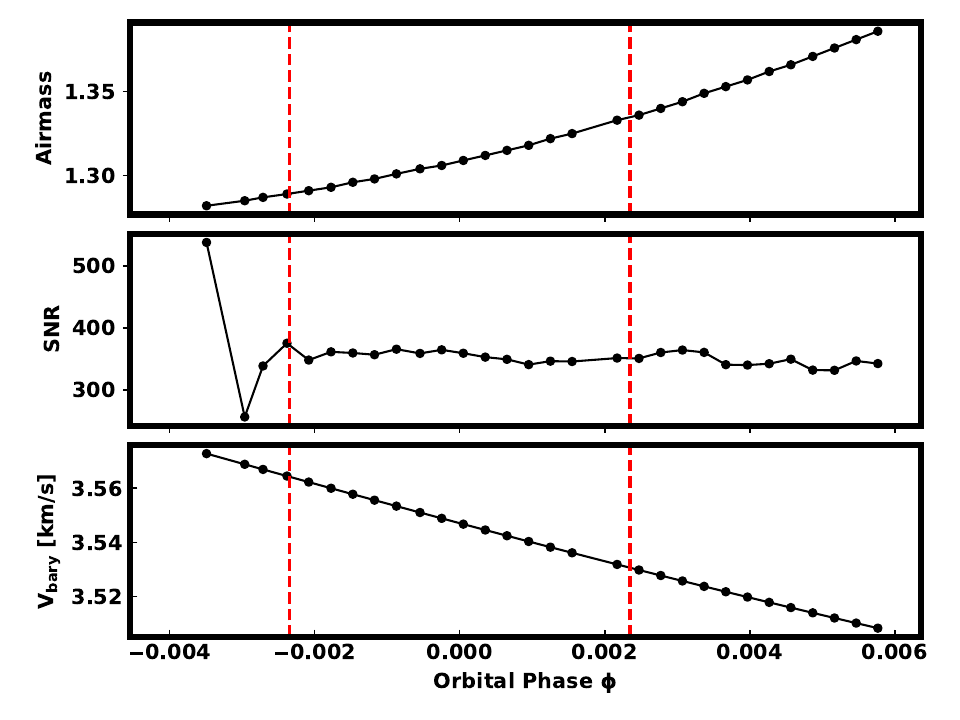}
    \raisebox{2.5em}{\includegraphics[width=0.49\linewidth]{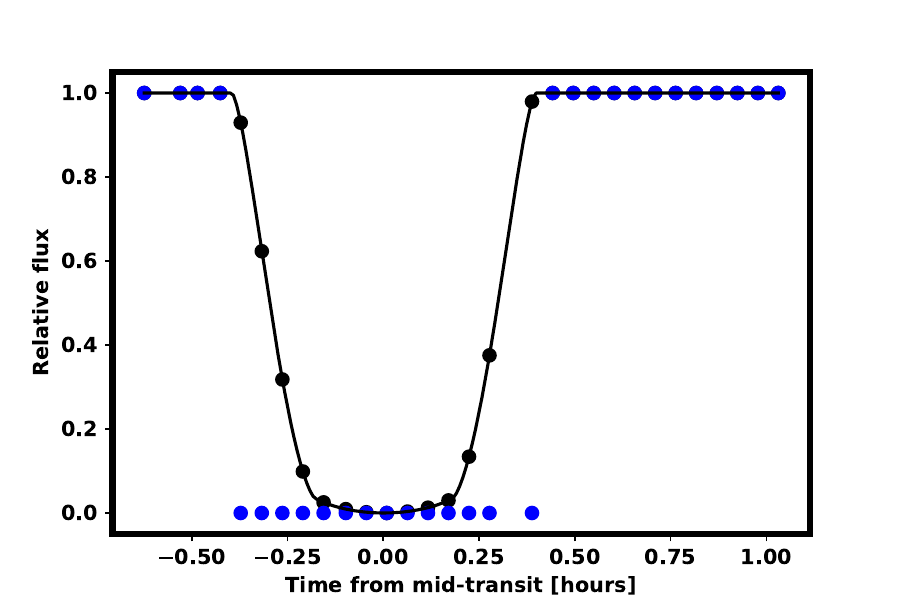}}
    \includegraphics[width=0.79\linewidth]{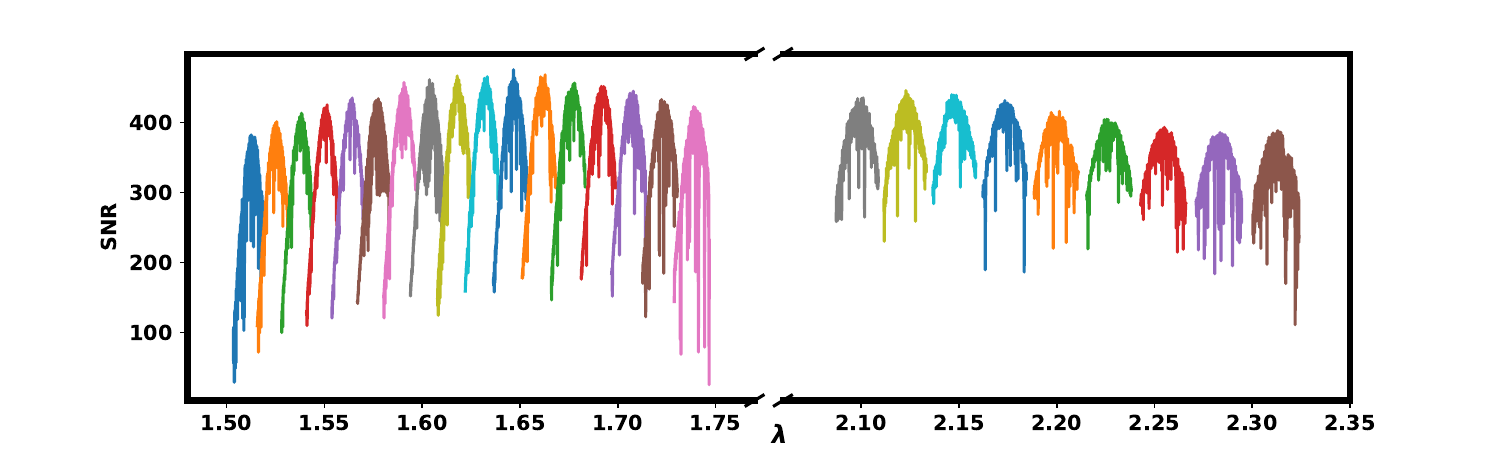}
    \caption{The observing conditions, data S/N, and lightcurve for the transit of L 98-59 d on the primary observing night. Left: we show the variation in airmass, the median S/N of the data in each exposure (after the removal of certain orders), and the changing barycentric velocity correction (positive defined as towards target) during the observations, as a function of orbital phase. The transit (determined using $T_{14}$; Table \ref{system_parameters}) is marked by the red dashed lines.
    Bottom: the median data S/N is now shown as a function of wavelength, with the different orders used in this work shaded.
    Right: the normalized lightcurve for L 98-59 d (black line), modelled using \texttt{batman-package} \citep{batman2015}. In this work, the time-series model spectra are modulated according to this profile during injection and recovery tests and likelihood-based model comparison. This modulation is shown for each of the exposures (black markers). Without this correction, the in-transit spectra are instead weighted equally (shown in blue), which may introduce biases during model comparison.}
    \label{fig:observing-conditions}
\end{figure*}

Spectra observed on both nights were taken in an A/B nodding configuration. For the night of observation on March 12 2021, this results in 30 AB nodding pairs throughout the night, which are combined to leave 30 background-subtracted spectra each with a total exposure time of 120\,s. The reduced spectra were provided by the instrument team at Gemini-S after this combination of the A/B frames. Table \ref{observing-info} lists the number of reduced spectra, the combined exposure time for each reduced spectrum (the total of the combined frames), and the total duration/phase range spanned by these reduced spectra, for each night of observation. This table also contains other information such as the median S/N of the data (after the removal of certain orders described in Section \ref{clean_norm}) and the split of in-transit/out-of-transit observations expected given the transit duration. Additionally, we present the S/N of the data, the variation in airmass, and the barycentric velocity correction during the primary night of observation in Figure \ref{fig:observing-conditions}, as a function of orbital phase.

\begin{table*}
\centering
\begin{tabular}{cccccccc}
\hline \hline
 Date & Obs. duration (hr) & Obs. phase range & Exp. time (s) & No. of spectra (in/out) & Airmass & Median S/N & $\Delta V_{\mathrm{p}}$\,($\mathrm{kms^{-1}}$) \\ \hline
 \hline
 10/02/2021 & 1.57 & -0.0032 -- 0.0056 & $\geq$240 & 13 (7/6) & 1.52 -- 1.88 & 620 & 1.83 \\ 
 \textbf{12/03/2021} & \textbf{1.65} & \textbf{-0.0035 -- 0.0058} & \textbf{120} & \textbf{30 (14/16)} & \textbf{1.28 -- 1.39} & \textbf{360} & \textbf{1.90} \\
 \hline
\end{tabular} 
\caption{Observational setup and conditions for each night of observation conducted under GO-Program GS-2021A-Q-216 (PI: P.R. McCullough). The information contained here refers to the spectra provided by the instrument team at Gemini-S, achieved after the combination of A/B nodding frames. The observed phase range and the number of in-transit spectra are calculated using parameters given in Table \ref{system_parameters}. 
The observation run on March 12 2021 used shorter exposure times (120\,s per A/B pair) than that on February 10 2021 ($\geq$240\,s per A/B combination), increasing the number of exposures taken throughout the observing night. We note that, for the observations of March 12 2021, these exposure times were 120\,s for each exposure after the first three (out-of-transit) exposures, for which the exposure times were variable (reflected in the S/N values achieved in Figure \ref{fig:observing-conditions}). 
Also note that the large uncertainties in the transit duration (Table \ref{system_parameters}) propagate into the calculation of the number of in-transit spectra (i.e. $14^{+3}_{-2}$), and motivate the modelling of the lightcurve during model comparison (Section \ref{lightcurve}). The change in the radial velocity of the planet during transit ($\Delta V_{\mathrm{p}}$) is calculated here from the difference between the first and last observed in-transit spectra (rather than the full transit duration), and includes the contribution from the changing barycentric velocity correction. The observing night primarily considered in this work is marked in bold, although an analysis of the second night is also conducted, with injection and recovery tests and cross-correlation results presented in Section \ref{appendix-night2}.}
\label{observing-info}
\end{table*}

\subsection{Change in radial velocity during transit}

In addition to its small expected atmospheric scale height, the longer period and shorter transit duration of L 98-59 d lead to a very small change in the radial velocity during transit. As a result, this planet is a challenging target for ground-based observations of this kind. Principal Component Analysis (PCA)-based detrending is motivated by the changing Doppler shift of the planetary spectral lines relative to the telluric and stellar lines. However, the change in radial velocity during transit is just $2.1\,\mathrm{kms^{-1}}$ in the absence of a barycentric velocity correction (this correction is included for the values given in Table \ref{observing-info}). 
To our knowledge, this is smaller than for any other exoplanet studied with ground-based high-resolution transmission spectroscopy to date, and
corresponds to the planet spectral lines crossing only $N_{\mathrm{px}}$ $\lesssim$ 1.0 instrument pixels during the transit. However, as demonstrated in \cite{cheverall2024}, high-resolution transmission spectroscopy may be feasible for similar temperate exoplanets when sufficient out-of-transit spectra are available during detrending. For each night of observation conducted under the current program (Table \ref{observing-info}), the ratio of the number of in-transit to out-of-transit spectra is $\sim$ 1, therefore encouraging the viability of these observations.

\subsection{Data cleaning, calibration, and normalization} \label{clean_norm}

The spectral data analyzed in this work were reduced, optimally extracted, and subjected to initial wavelength calibrations by the IGRINS Pipeline Package \citep[\texttt{PLP};][]{sim2014, mace2018} and were provided by the instrument team at Gemini-S. Throughout this work we analyze the reduced time-series spectra which have undergone initial telluric correction by the \texttt{PLP} using an A0V telluric standard star and a spectrum for Vega. We find that this telluric correction is not sufficient, and the residuals still require further correction in order to access the planet signal, as described in Section \ref{telluric-correction-methods}. 
However, this initial correction step may be helpful as it uses a real measurement of the telluric spectrum on the night of observation, and provides some initial telluric correction of spectra before the embedded planet signal is degraded through subsequent application of PCA.
A comparison of our results with and without this initial telluric correction setting can be seen later in this manuscript (Section \ref{considerations}).

After visual inspection of the spectral data, we remove 27 orders (0-5, 23-37, 47-52 inclusive) from our data set due to poor data quality/increased telluric contamination. This is the same for both observing nights, leaving 26 spectral orders on each night for the remainder of this work. We additionally trim low-throughput pixels at the edges of each remaining spectral order, similar to e.g. \cite{brogi2023, smith2024_igrins, kanumalla2024}. We follow the methods laid out in \cite{cheverall2024} to clean and normalize the reduced spectra. Treating each spectral order individually, the reduced spectra are rescaled, cleaned of bad pixels and outliers, and then normalized using a second-order polynomial fit to the continuum of each order and exposure.

Deep and broad molecular features originating from the host M-dwarf star may be difficult to remove from the continuum with a polynomial fit, and these may lead to some distortion and modulation of the planetary spectra. However, this may be somewhat mitigated by model reprocessing (Section \ref{bayesian-methods}). We also note that molecules abundant within the stellar photosphere have the potential to imprint spurious phase-correlated signatures onto planetary transmission spectra through the Rossiter-McLaughlin effect \citep[e.g.][]{rossiter_detection_1924, mclaughlin1924, queloz_detection_2000}. However, with the exception of \ce{H2O}, the molecules we consider here are unlikely to be present/abundant in the stellar photosphere, and therefore should not be susceptible to such false-positives.

Following previous works \citep[e.g.][]{line2021, brogi2023, smith2024_igrins}, we perform a secondary wavelength calibration to account for any drift in the wavelength solution during the course of the observations. To do this, we apply a second-order stretch and shift transform to the wavelength grid of each spectrum, such that it aligns with the spectrum at the end of the observing night, which was taken closest in time to the spectrum used for the initial wavelength calibration performed by the \texttt{PLP}. In our case, we find that the required correction is minimal in comparison to similar works (less than $\pm0.1$ pixels or 0.2\,$\mathrm{km\,s^{-1}}$).

\subsection{Telluric correction and cross-correlation} \label{telluric-correction-methods}

In order to extract the planetary spectra, the observed time-series spectra (with or without the initial telluric correction provided by IGRINS \texttt{PLP}) must be corrected for the telluric and stellar contributions, which conceal the planet signal. The telluric contamination is particularly severe and variable in the NIR, and so a data-driven approach is often required for telluric correction; PCA is commonly used for this purpose \citep[e.g.][]{de_kok_detection_2013, giacobbe_five_2021, lafarga2023}. In this work, we use the $\Delta$CCF framework \citep[e.g.][]{holmberg_first_2022, spring_black_2022, cheverall_robustness} to determine an optimum number of principal components to subtract during detrending, so as to not systematically bias the detection significance by overfitting detrending parameters \citep{cheverall_robustness}. This offers a more robust approach than optimising the detection S/N directly, which is prone to overfitting noise and therefore leading to biased signal recoveries \citep[e.g.][]{cabot_robustness_2019, cheverall_robustness}. 
Using the $\Delta$CCF framework, we calculate the optimum number of principal components to subtract for each of the molecules considered in this work. Subtracting a different number of PCA iterations for each molecule considered is suitable in our present study, where we consider each species one at a time and do not perform a joint retrieval of the chemical abundances across species.
For \ce{H2O}, we enforce that the minimum number of components to be subtracted is 2 in order to ensure sufficient telluric removal, following \cite{cheverall_robustness}. We note that, in Section \ref{diversity}, whilst the optimum number of principal components to subtract may vary slightly depending on the atmospheric conditions modelled (parameterised by chemical abundance and cloud-top pressure), we continue to subtract the number of components determined as the optimum for the nominal model of that molecule, as defined in Section \ref{modelling}.

\begin{table}
\centering
\begin{tabular}{cc}
\hline \hline
 Model Species & Principal Components \\ \hline
 \hline
 \ce{CH4} & 7 \\
 \ce{CO2} & 3 \\
 \ce{CO} & 2 \\
 \ce{NH3} & 1 \\
 \ce{H2S} & 1 \\
 \ce{H2O} & 4 \\
 \hline
\end{tabular} 
\caption{The optimum number of principal components to subtract during detrending for the nominal models of each of the molecules considered in this work, obtained through the $\Delta$CCF framework. For the species considered in Section \ref{diversity}, where the chemical abundance and atmospheric cloud-top pressure are varied, we continue to subtract the number of components determined using the nominal model (Section \ref{modelling}), for consistency.
}
\label{delta-ccf-table}
\end{table}

After detrending, the spectral residuals are cross-correlated with Doppler-shifted model spectra for each of the chemical species considered (further details for these models can be found in Section \ref{modelling}). As has been well-established \citep[e.g.][]{brogi_signature_2012}, cross-correlation is able to combine the information from all individual planet spectral lines (each with S/N $<<$ 1) into a measurable signal \citep[e.g.][]{birkby_exoplanet_2018}. For a chemical detection of a given species, a cross-correlation peak will be seen when the model is Doppler-shifted by the expected velocity of the planet. In this work we calculate a cross-correlation function (CCF) as a function of orbital phase and radial velocity, between $\pm400$\,$\mathrm{kms^{-1}}$ in steps of 0.1\,$\mathrm{kms^{-1}}$. Considering the trail of a planet through this CCF, a S/N is then calculated as a function of the radial velocity semi-amplitude of the planet $K_{\mathrm{p}}$ and the systemic velocity of the star $V_{\mathrm{sys}}$, accounting for the known barycentric velocity correction at the time of each observation ($V_{\mathrm{bary}}(t)$; Figure \ref{fig:observing-conditions}). In this S/N calculation, the signal is taken as the sum of CCF values along a given planetary trail, whilst the noise is estimated using the standard deviation of CCF values away from this trail (in this work more than 15\,$\mathrm{kms^{-1}}$ away). We calculate the S/N across a grid given by $K_{\mathrm{p}}$ of 10--150\,$\mathrm{kms^{-1}}$ and $V_{\mathrm{sys}}$ of $\pm40$\,$\mathrm{kms^{-1}}$.

In this work we use cross-correlation only to perform an initial, first-pass assessment of the data. For simplicity, we therefore do not weight the time-series spectra in the cross-correlation function according to the lightcurve of the planet, nor reprocess the time-series model to replicate onto it the effects of PCA-based detrending, as is done in the more comprehensive likelihood-based approach (Section \ref{bayesian-methods}). We do however weight the cross-correlation function according to the inverse square of the error in each wavelength channel \citep[e.g.][]{gibson2020, nortmann2024, cont2025}, estimated by the median absolute deviation of the time-series residuals.

\subsection{High-resolution model spectra} \label{modelling}

Model atmospheric transmission spectra are generated for L 98-59 d in a similar manner to the models in \cite{cheverall2024}, using a variant of the AURA atmospheric modelling code for exoplanets \citep{pinhas_retrieval_2018}. Assuming a plane-parallel \ce{H2}-rich atmosphere in hydrostatic equilibrium and an isothermal temperature profile, line-by-line radiative transfer is computed across the slant path length over a pressure range of $10^{-7}$ - $100$ bar. 

\begin{figure*}
    \centering
    \includegraphics[width=0.8\textwidth]{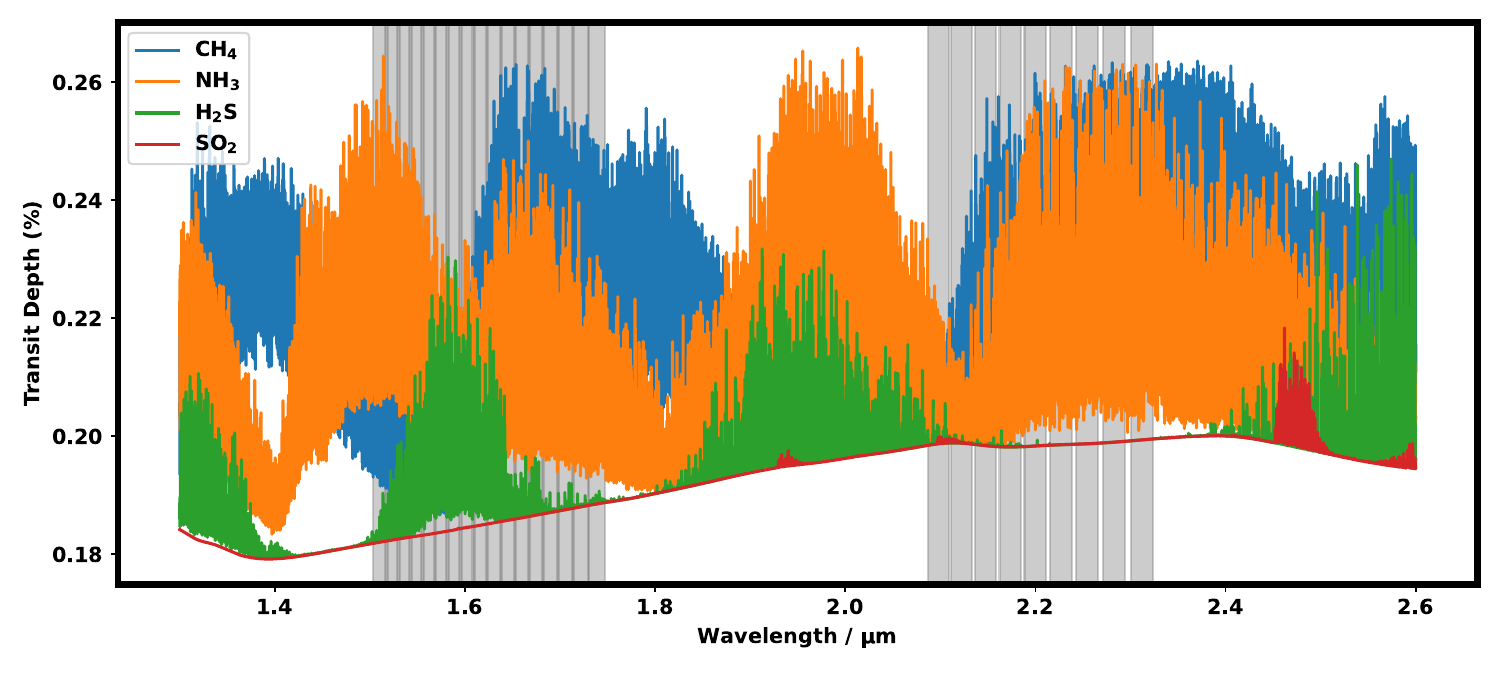}
    \caption{Model spectra for a selection of the molecular species considered in this work. \ce{CH4} and \ce{NH3} are expected to be prevalent and detectable in temperate \ce{H2}-dominated atmospheres using this technique \citep{cheverall2024}, whilst \ce{H2S} and \ce{SO2} have been tentatively inferred previously in the atmosphere of L 98-59 d \citep{banerjee2024}. In addition to the molecules shown here, opacity contributions from \ce{CO2}, \ce{CO}, and \ce{H2O} are considered. The spectral orders used in this analysis remaining after initial data cleaning are marked in gray.}
    \label{fig:models}
\end{figure*}

High-resolution spectra are generated for one molecular species at a time, spanning the IGRINS spectral range in the NIR. Opacity contributions from several molecules which may be expected in this particular temperate \ce{H2}-dominated atmosphere are considered (\ce{CH4}, \ce{NH3}, \ce{H2S}, \ce{SO2}, \ce{CO}, \ce{CO2}, and \ce{H2O}). The nominal models used initially in Section \ref{results} (full model grids are later explored in Section \ref{diversity}) assume no clouds/hazes and a nominal temperature of 300\,K, close to the equilibrium temperature of L 98-59 d. Nominal mixing ratios of $5\times10^{-3}$, $10^{-3}$, $10^{-3}$, $10^{-3}$, $3\times10^{-4}$, $10^{-4}$, and $10^{-2}$ are assumed for \ce{CH4}, \ce{NH3}, \ce{CO2}, \ce{CO}, \ce{H2S}, \ce{SO2}, and \ce{H2O} respectively. These nominal mixing ratios correspond to metallicities of approximately 10$\times$ solar, and are chosen to be consistent with previous observations of comparable sub-Neptune exoplanets \citep[e.g.][]{madhusudan_carbon}. A selection of these models are plotted in Figure \ref{fig:models}. Molecular cross-sections were obtained \citep[see][]{gandhi_genesis_2017} using the following absorption line lists: H$_2$O \citep{Polyansky2018}, CH$_4$ \citep{yurchenko_exomol_2014}, and NH$_3$ \citep{yurchenko_variationally_2011}, \ce{CO} \citep{Li2015}, CO$_2$ \citep{Tashkun2015}, SO$_2$ \citep{Underwood2016} and \ce{H2S} \citep{azzam2016, chubb2018}. Collision-induced absorption from H$_2$-H$_2$ and H$_2$-He was also included in modelling \citep{borysow1988,orton2007,abel2011,richard_new_2012}. The model spectra are separated into orders and normalized in the same way as the spectral data, and convolved with the point spread function of the instrument, before comparison with the data.

In Section \ref{diversity}, we vary the chemical abundance and cloud-top pressure parameters to generate a grid of atmospheric models over which we can explore. We model the spectral contributions of atmospheric clouds using a grey cloud model parametrised by the cloud-top pressure, assuming a full coverage of the terminator atmosphere. Lower cloud-top pressures result in spectral features becoming truncated from below due to masking from cloud absorption, leaving fewer lines available for atmospheric characterisation.

\subsection{Considerations for Orbital Geometry} \label{lightcurve}

The large impact parameter of L 98-59 d ($b=0.922\pm0.059$) gives rise to a significant difference between $T_{14}$ and $T_{23}$ and significant uncertainties associated with each (Table \ref{system_parameters}). This corresponds to the lightcurve of L 98-59 d being more sloped than for planets with low impact parameters (Figure \ref{fig:observing-conditions}), for which the transit limbs are steeper and the difference between $T_{14}$ and $T_{32}$ is reduced. Therefore, for the model injection/recovery tests in this work, and during the likelihood-based model comparison introduced in the following section, we modulate the transit depths of the atmospheric models from the previous section (generated assuming a flat stellar disk) in time accordingly, to account for the expected profile of the lightcurve in the model time-series spectra. To do this, we model the lightcurve using \texttt{batman-package} \citep{batman2015} and the limb darkening coefficients given in Table \ref{system_parameters} \citep[\texttt{exoctk;}][]{bourque_2021}.  Although it is difficult to constrain the limb-darkening parameters given the high impact parameter \citep{gressier2024}, their precision is sufficient for our current use (weighting the time-series model spectra) and the associated uncertainties have minimal impact on the final detection significance reported in Section \ref{results}. Without this correction (instead weighting model spectra equally across all in-transit spectra as is commonly done; see blue points in Figure \ref{fig:observing-conditions}), the analysis is more vulnerable to uncertainties in $T_{14}$, and it is possible that injected model spectra will be too strong, particularly at ingress/egress, introducing false inferences of detectability during injection and recovery tests, or biases in the model scale factor $\alpha$ during model comparison (Section \ref{bayesian-methods}). For example, the total scale factor of the injected model signal, integrated over the transit, is increased by $\sim$ 34\,$\%$ for this dataset when injecting equally into all in-transit spectra, compared to when using the weightings described here.

Despite this, when calculating the CCF (and subsequently the cross-correlation S/N) in the cross-correlation framework, we note that the time-series spectra and model are not weighted according to the planet lightcurve\footnote{However, we do apply a weighting to this calculation according to the errors in each wavelength channel, as described in Section \ref{telluric-correction-methods}.}, in order to pursue a first-pass assessment of the data agnostic to the orbital parameters of the planet.

\subsection{Model reprocessing and comparison} \label{bayesian-methods}

An alternative to cross-correlation is to consider the probabilistic likelihood of each atmospheric model, building upon various works from across the literature \citep[e.g.][]{brogi_retrieving_2019, gandhi_hydra-h_2019, gibson2020, gibson_relative_2022, lafarga2023, brogi2023}. We here describe the methods used in this work, including the model reprocessing, the model comparison with the spectral residuals, and finally the calculation of a detection significance for each model.

After cleaning, normalizing, and detrending the spectra as before, the spectral residuals are compared with a reprocessed model template. The model is reprocessed so as to replicate onto the model the effects of PCA-based detrending on any real planet spectrum in the data, to ensure a like-for-like comparison. We perform this model reprocessing as follows. First, we empirically derive a model telluric/stellar template from the data by summing the principal components of the data which were subtracted during detrending (the initial spectra before detrending minus the detrended residuals). Next, we inject the Doppler-shifted and scaled time-series planet model spectrum into this telluric/stellar template, to simulate pre-detrended spectra which contain this specific planet model spectrum. In this step, the atmospheric model spectra are Doppler shifted by a planetary velocity, parameterised by $K_{\mathrm{p}}$ and $V_{\mathrm{sys}}$, modulated by the expected lightcurve of the planet (Section \ref{lightcurve}), and then multiplied by an atmospheric scaling factor $\alpha$. This scaling factor makes the likelihood fit sensitive to line depths \citep{brogi_retrieving_2019} in addition to the relative depths and positions of spectral lines, and may help to distinguish real planet signals ($\alpha \approx 1$ for the correct model) from spurious signals. Finally, we detrend these simulated spectra in the same way as we did for the data, to replicate the effects of the PCA-based detrending onto the time-series model spectra. In addition to reproducing the erosion and distortion of planetary spectra resulting from detrending, this step may also replicate onto the model any modulation of the planetary spectra arising from stellar absorption/activity or instrumental effects. By giving a closer match between any residual planet signal and the model template, such model reprocessing may help to avoid detection biases \citep{brogi_retrieving_2019} and increase detection sensitivity.

After this reprocessing, the time-series atmospheric model $m_{ij}$ (with model parameters $\theta$: $K_{\mathrm{p}}$, $V_{\mathrm{sys}}$, and $\alpha$) can be compared to the spectral residuals $r_{ij}$ (across wavelength channel $j$ and observation $i$) using Equation \ref{logl_calc}. The model likelihood calculated gives the probability of the data given the atmospheric model, as function of the model parameter space. In this work, the error in each wavelength channel of the residuals $\sigma_j$ is assumed to be independent of time, and is estimated by the median absolute deviation of the detrended residuals in that wavelength channel.
\begin{equation} \label{logl_calc}
\mathrm{log}(L)(\theta) = -\frac{1}{2} \sum_{i, j}\left[\left(\frac{r_{ij} - m_{ij}(\theta)}{\sigma_{j}}\right)^2 + \mathrm{log}(2\pi\sigma_{j}^2)\right] 
\end{equation}
Using Equation \ref{logl_calc}, we can calculate a posterior probability distribution for each atmospheric model using \texttt{emcee} \citep{emcee2013}, assuming the prior probability distributions for model parameters $\pi(\theta)$ given in Table \ref{priors} (e.g. Figure \ref{fig:test}).

\begin{table}
\centering
\begin{tabular}{ccc}
\hline \hline
 Model Parameter & Expected Value & Prior Range\\ \hline
 \hline
 $K_{\mathrm{p}}$ & 70\,$\mathrm{kms^{-1}}$ & 40 -- 100\,$\mathrm{kms^{-1}}$ ($\pm$30\,$\mathrm{kms^{-1}}$) \\ 
 $V_{\mathrm{sys}}$ & -5.6\,$\mathrm{kms^{-1}}$ & -15.6 -- 4.4\,$\mathrm{kms^{-1}}$ ($\pm$10\,$\mathrm{kms^{-1}}$) \\ 
 $\alpha$ & 1.0 & 0.3 - 3.0 \\
 \hline
\end{tabular} 
\caption{Prior probability distributions for the model parameters.}
\label{priors}
\end{table}

\subsection{Computation of detection significance} \label{bayes-calculation}

In addition to this posterior probability distribution, we additionally calculate the Bayesian model evidence for each atmospheric model considered, which acts as a measure of the posterior probability of that atmospheric model. To calculate this model evidence, we marginalize the likelihood distribution over the model parameters, modulated by their prior probability distributions (Table \ref{priors}) as follows.

\begin{equation} \label{evidence_calc}
Z_{m} = \int_{\theta} L\pi(\theta) d\theta
\end{equation}

For each atmospheric model of a given chemical species, we compare the model evidence to the evidence calculated for a null atmospheric model not containing that species (in this case a flat line, corresponding to the model with $\alpha$ = 0). For this comparison, we compute the Bayes factor $B$ for each atmospheric model as the ratio of these model evidences ($Z_{\mathrm{model}}/Z_{\mathrm{null}}$). This Bayes factor gives the ratio of the posterior probability of the atmospheric model to that of the null model (i.e. the odds ratio in favor of an atmospheric model containing a particular chemical species), and therefore acts as a strong measure of confidence for the possible presence of a given species. To aid interpretation, the Bayes factor is often converted to an estimate of the frequentist detection significance $\sigma$ for each model \citep[e.g.][]{sellke2001, benneke2013}. In this work, we perform this estimation of the detection significance of the atmospheric model relative to the null model using the expression given in Equation 17 of \cite{welbanks2021}. We note that this conversion is an estimate only, providing an upper bound on the frequentist detection significance \citep[see e.g.][]{kipping2025, thorngren2025}. As a result, for our marginal chemical inference presented in Section \ref{results} and in our exploration of model grids in Section \ref{diversity}, we additionally report the Bayes factor alongside the converted detection significance.

\section{Results} \label{results}

In this section we use the methods outlined previously to analyze the high-resolution IGRINS spectra described in Section \ref{observations} and search for spectral signatures of several common molecules. We primarily consider \ce{CH4} and \ce{NH3}, which have been predicted to be easily detectable in \ce{H2}-rich atmospheres of temperate sub-Neptune planets, such as L 98-59 d, using high-resolution transmission spectroscopy \citep{cheverall2024}. We additionally consider \ce{H2S} and \ce{SO2}, given the previous inference for the presence of these molecules in the atmosphere of L 98-59 d \citep{gressier2024, banerjee2024}, as well as \ce{CO}, \ce{CO2}, and \ce{H2O} which may be expected in an atmosphere of this kind. The following results demonstrate the feasibility of characterizing the atmospheric molecular diversity of small, temperate super-Earths, such as L 98-59 d, using high-resolution transmission spectroscopy with 8\,m-class ground-based telescopes.

\subsection{Initial characterization of atmospheric composition} \label{initial-cross-correlation}
 
\begin{figure*}
    \centering
    \includegraphics[width=\textwidth]{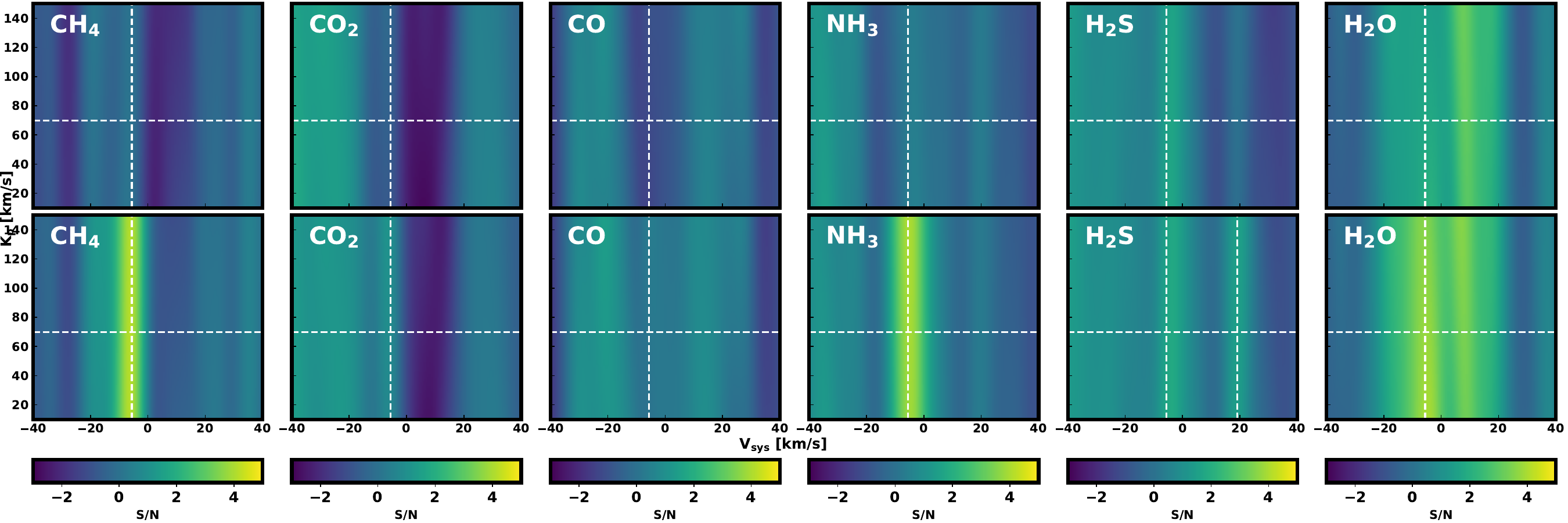}
    \caption{Cross-correlation results for the observed time-series spectra and models of the molecular species considered in this work. The cross-correlation S/N is shown as a function of planetary velocity space, parameterised by $K_{\mathrm{p}}$ and $V_{\mathrm{sys}}$, for different molecular species. In each case, the expected planetary velocity is marked by white crosshairs. For consistency, the colorscale is fixed to [-3, 5] across all panels. Top panel: when cross-correlating atmospheric models with the observed data, no significant signals are seen, although a tentative peak of $\mathrm{S/N} = 1.7$ is seen for \ce{H2S} consistent with the expected $V_{\mathrm{sys}}$. Bottom panel: injection and recovery tests are used to assess detectability, with a model atmospheric spectrum for each chemical species injected into the spectra prior to detrending. This injection is at the expected systematic velocity, except in the case of \ce{H2S}, where the signal is injected at an alternative systematic velocity ($+$19\,$\mathrm{kms^{-1}}$, chosen as a region of velocity space with an initial S/N close to 0, and marked by a second crosshairs) in order to avoid adding to the observed peak. Recovered signals at the velocities of injection suggest that our analysis is moderately sensitive to \ce{CH4} and \ce{NH3}, with additional limited sensitivity to \ce{H2O} (also see Figure \ref{fig:1d-profiles} for 1-D cross-sections at the expected value for $K_{\mathrm{p}}$). 
    }
    \label{fig:detection-survey_primary}
\end{figure*}

To identify in the spectral residuals any signatures of the molecular species considered here (\ce{CH4}, \ce{CO2}, \ce{CO}, \ce{NH3}, \ce{H2S}, \ce{H2O} and \ce{SO2}), we initially use the cross-correlation framework described in Section \ref{methods}. In Figure \ref{fig:detection-survey_primary}, the cross-correlation S/N is shown across planetary velocity space for each molecule. To first demonstrate the feasibility of obtaining a chemical inference of each molecule, injection and recovery tests are conducted in the bottom panel. 
We inject the nominal atmospheric models at 1$\times$ strength (Section \ref{modelling}) scaled by the lightcurve described in Section \ref{lightcurve} to simulate a single transit, with successful recoveries for a sample of molecules indicating that, using this technique, it is indeed possible to achieve significant chemical inferences in the atmospheres of such planets. We find that we are moderately sensitive to \ce{CH4} and \ce{NH3} (for which we recover the injected signals with S/N values of 4.1 and 4.0, respectively), with additional limited sensitivity to \ce{H2O}. We confirm that our analysis is not at all sensitive to the presence of \ce{SO2} due to the lack of available spectral features in this wavelength range (Figure \ref{fig:models}) and therefore we do not show it here or consider it further in this work. Results from these injection and recovery tests are further shown as one-dimensional S/N profiles in Figure \ref{fig:1d-profiles}. In the top panel of Figure \ref{fig:detection-survey_primary}, we present cross-correlation results for the observed data without any model injections. 
No strong cross-correlation peaks are observed at the expected systemic velocity for the molecules considered here. However, we do observe a tentative cross-correlation peak for \ce{H2S} at the expected systemic velocity. Whilst this peak is of relatively low S/N (1.7), and therefore not strong enough to alone constitute a chemical detection of this species, its presence motivates the injection of the planet signal during testing into an area of velocity space where we initially find a S/N near 0. This is in order to avoid the addition of the injected signal to the observed peak, which would overestimate the sensitivity to this molecular species. In this case, we select a systematic velocity of $+$19$\,\mathrm{kms^{-1}}$ for this purpose, as shown in the bottom panel of Figure \ref{fig:detection-survey_primary}. Given the S/N achieved in this injection and recovery test (1.6), we note that the S/N of the aforementioned signal in the data is consistent in both velocity and strength with what would be expected for a chemical detection of \ce{H2S} in the atmosphere of this planet. Whilst the S/N is low, 
this consistency motivates further exploration of this cross-correlation signal in Section \ref{h2s-follow-up} and throughout this work.

\subsection{A tentative \ce{H2S} signal: further analysis} \label{h2s-follow-up}

In this section, we further investigate the tentative cross-correlation peak observed for \ce{H2S}. As discussed in \cite{cheverall2024}, the radial velocity semi-amplitude $K_{\mathrm{p}}$ cannot typically be constrained from the slope of a cross-correlation signal for planets which undergo only a small change in radial velocity during transit, such as L 98-59 d. However, in such cases, we may instead measure the duration of a cross-correlation signal in time. A signal arising from the absorption of a species in the atmosphere of L 98-59 d should be constrained to only the transit duration of the planet, providing a test of its origin and helping to distinguish it from spurious stellar signals. Following the method outlined in \cite{cheverall2024}, we perform this test for the set of molecular species considered in this work.

\begin{figure*}
    \centering
    {\includegraphics[width=0.49\textwidth]{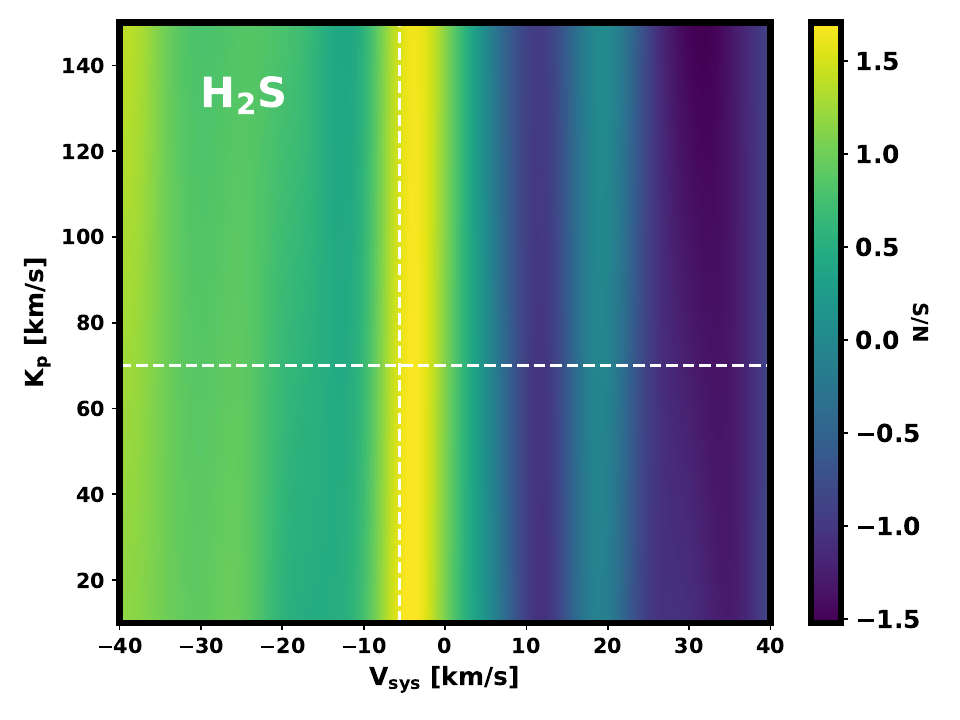}}
    {\includegraphics[width=0.49\textwidth]{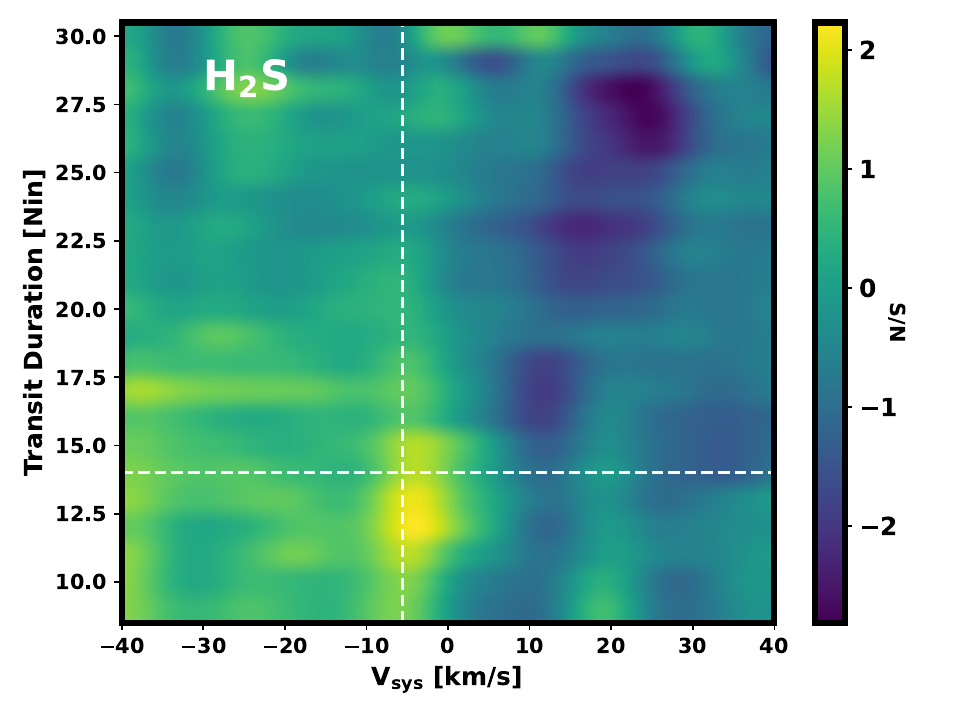}}
    \caption{Additional analysis for the tentative cross-correlation signal observed for \ce{H2S}. We show the cross-correlation S/N as a function of the orbital parameters of the planet.
    In each case, the expected parameters are marked with crosshairs. 
    Left: The S/N is shown as a function of planetary velocity space, parameterised by $K_{\mathrm{p}}$ and $V_{\mathrm{sys}}$, reproduced from the top panel of Figure \ref{fig:detection-survey_primary} but with a more zoomed in colorscale. A tentative peak with a S/N of 1.7 is observed at the expected systemic velocity. 
    Right: the time duration of the observed \ce{H2S} cross-correlation signal is constrained with a value that is consistent with the literature value for the transit duration of the planet. This is measured by varying the number of spectra considered to be in-transit ($N_{\mathrm{in}}$) when calculating the cross-correlation S/N, with $K_{\mathrm{p}}$ fixed to that expected for this planet. 
    }
    \label{fig:e-follow-up}
\end{figure*}

Our findings for \ce{H2S} are shown in the left-hand panel of Figure \ref{fig:e-follow-up}. We vary on the y-axis the number of spectra considered to be in-transit ($N_{\mathrm{in}}$) when calculating the cross-correlation S/N (the S/N is calculated from only the in-transit spectra), whilst fixing $K_{\mathrm{p}}$ to that expected for this planet (70\,$\mathrm{kms^{-1}}$). We find that the duration of the signal is indeed constrained and consistent with the literature value of the transit duration of the planet (which corresponds to $14^{+3}_{-2}$ in-transit spectra). 
We note that in this test we are weighting the "in-transit" spectra equally, and so for planets with large impact parameters, such as current target, we are approximating the equivalent transit duration of a box-shaped lightcurve, rather than the transit duration of the planet itself.
The slightly smaller than literature value found here is therefore consistent with what would be expected given the small transit depths of in-transit spectra at ingress/egress in Figure \ref{fig:observing-conditions}. As a result, we may have increased confidence that the observed cross-correlation peak could be planetary in origin, through independently measuring the duration of the signal (despite the coarse sampling in time) to be consistent with the literature value for the transit duration of the planet. Now using the number of in-transit spectra for which the cross-correlation S/N is at its maximum (12), the recalculated cross-correlation result is shown in Figure \ref{fig:optimum-nin}, with a S/N of 2.3. Meanwhile, we observe no constrained signals anywhere in this space when cross-correlating with the other molecules considered in this work.

Following this, we also perform a number of cross-correlation robustness tests. First, we check that the recovered planet signal follows the same characteristic behavior predicted for low-velocity planets in \cite{cheverall2024}, by considering some tests proposed in that work. For example, we confirm that the signal is indeed lost when the out-of-transit spectra are removed prior to detrending (despite no change to the in-transit spectra used to calculate the S/N), consistent with the demonstrated importance of out-of-transit spectra for successful detrending for temperate exoplanets. Additionally, we cross-correlate the model with just the out-of-transit spectra after detrending. Whilst this would remove the signal originating from any transiting planet, it was simulated previously that for low-velocity planets, an anti-correlation peak should be found in the out-of-transit CCF, as an artifact of the detrending. In this current work, we indeed find such an anti-correlation peak ($\mathrm{S/N}=-1.8$) at the expected systematic velocity. Second, we verify that the spectral lines of different molecular species do not cross-correlate with each other to give spurious detections. For example, we do not observe cross-correlation signals when injecting spectra for different molecules and then cross-correlating with \ce{H2S}, demonstrating that degeneracy between species is broken at the high spectral resolutions used here.

\subsection{Likelihood-based framework}

\begin{figure}
    \centering
    \includegraphics[width=0.49\textwidth]{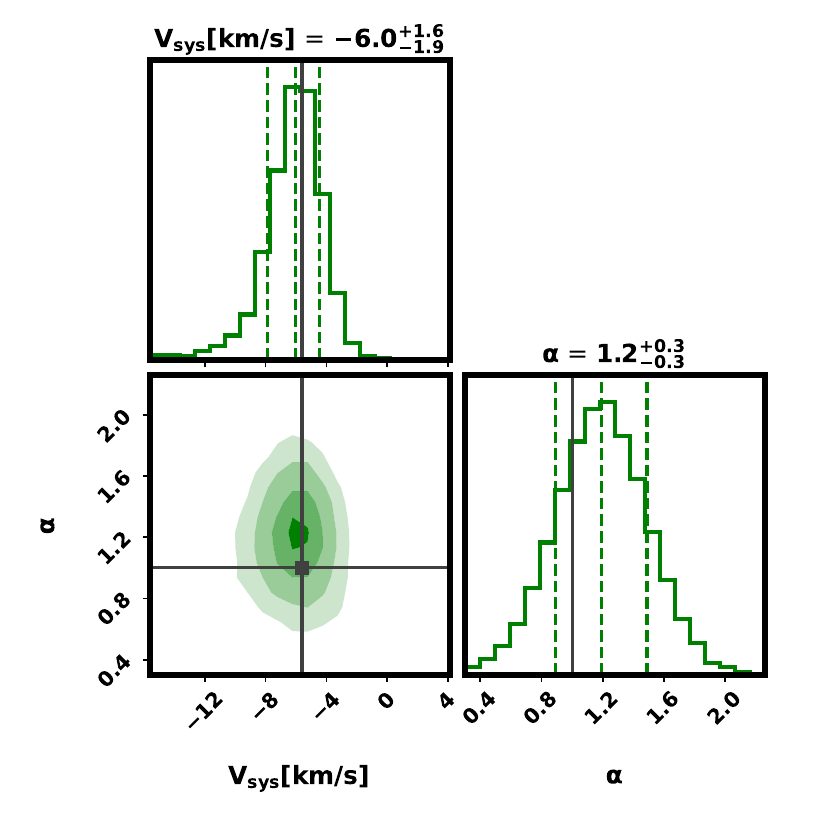}
    \caption{The posterior probability distribution for \ce{H2S} as a function of the planetary parameters. The expected parameters are marked with crosshairs. The values for $V_{\mathrm{sys}}$ and $\alpha$ measured from the posterior probability distribution are consistent with those expected. The 0.5$\sigma$, 1.0$\sigma$, 1.5$\sigma$, and 2$\sigma$ contours are shown in the 2D-plot, with the 1$\sigma$ bounds also shown in each 1D histogram. For temperate planets such as L 98-59 d, with small changes in radial velocity during transit, it is difficult to constrain $K_{\mathrm{p}}$ and therefore it is not included as a panel in this corner plot.}
    \label{fig:test}
\end{figure}

Following the methodology outlined in Section \ref{bayesian-methods}, we now instead use a likelihood framework to compare the detrended spectral residuals with a model template for each molecule. In doing so, we calculate the posterior probability distribution for the presence of each molecule in the atmosphere of L 98-59 d (corner plot; Figure \ref{fig:test}). For \ce{H2S}, clear preference is shown for an atmospheric model with systematic velocity and scale factor parameters (measured to be $V_{\mathrm{sys}}$ = $-6.0^{+1.6}_{-1.9}$\,$\mathrm{kms^{-1}}$ and $\alpha$ = $1.2\pm0.3$) which are consistent with the expected values ($V_{\mathrm{sys}}$ = $-5.6$\,$\mathrm{kms^{-1}}$ and $\alpha =$ 1.0), adding further support to a chemical detection. Small discrepancies in the model scale factor $\alpha$ from 1.0 can be attributed to the choice of nominal atmospheric model considered here, e.g. the temperature and chemical abundance used, whilst there is not the required sensitivity or precision on the measured $V_{\mathrm{sys}}$ to infer the presence of any winds at the terminator. No signals consistent with a planetary origin are seen in the posterior probability distributions for the other molecules considered. Additionally, as described in Section \ref{bayes-calculation}, we marginalize the likelihood distribution over model parameter space (the priors for the model parameters are given in Table \ref{priors}). A Bayes factor of 390 is calculated in preference of the nominal \ce{H2S} atmospheric model over the null model, exceeding the threshold typically considered to indicate a confident detection. Using the relevant conversion to a frequentist measure of significance, this corresponds to a frequentist detection significance of 3.9$\sigma$. Meanwhile, a Bayes factor of $>$0.5 is not found for any of the other molecules considered in this work.

For the inference of \ce{H2S}, we note the increased significance compared to what was found using the cross-correlation framework. To investigate this, we now repeat and compare the injection and recovery tests from Section 3.1 using the likelihood-based framework, for identical injections into the same data. For example, whereas in Figure \ref{fig:detection-survey_primary} S/N values of 4.0 and 4.1 are found for \ce{NH3} and \ce{CH4}, we here recover identical injections with significances of 6.8$\sigma$ and 6.1$\sigma$. For \ce{H2S}, we recover the same injected signal as previously (at a systemic velocity of 19\,$\mathrm{kms^{-1}}$) to 4.6$\sigma$, compared to the cross-correlation S/N of 1.5 (Section \ref{initial-cross-correlation}). The present estimation of the Bayesian detection significance in this work is therefore empirically shown to be more sensitive than the cross-correlation S/N, for the data and models considered here. This may owe to the improved handling of errors, the model reprocessing to account for the degrading effects of PCA, and the consideration of the lightcurve using this framework. Whilst these improvements are indeed expected to increase the sensitivity of the likelihood approach compared to cross-correlation (and this is reflected in the calculated Bayes factors), we also note that the Bayesian detection significance values calculated here are upper-bound estimates, as described in Section \ref{bayes-calculation}. As a result, we also quote the calculated Bayes factor when discussing significances above and in the remainder of this work.

\subsection{Robustness against pipeline settings} \label{considerations}

To provide further robustness and assess the contribution of each part of the methodology to the final result (and the associated variation in detection significance introduced by each), we now reproduce our results for \ce{H2S} whilst varying our implementation at different stages in the analysis.

\begin{figure*}
    \centering
    \includegraphics[width=0.33\textwidth]{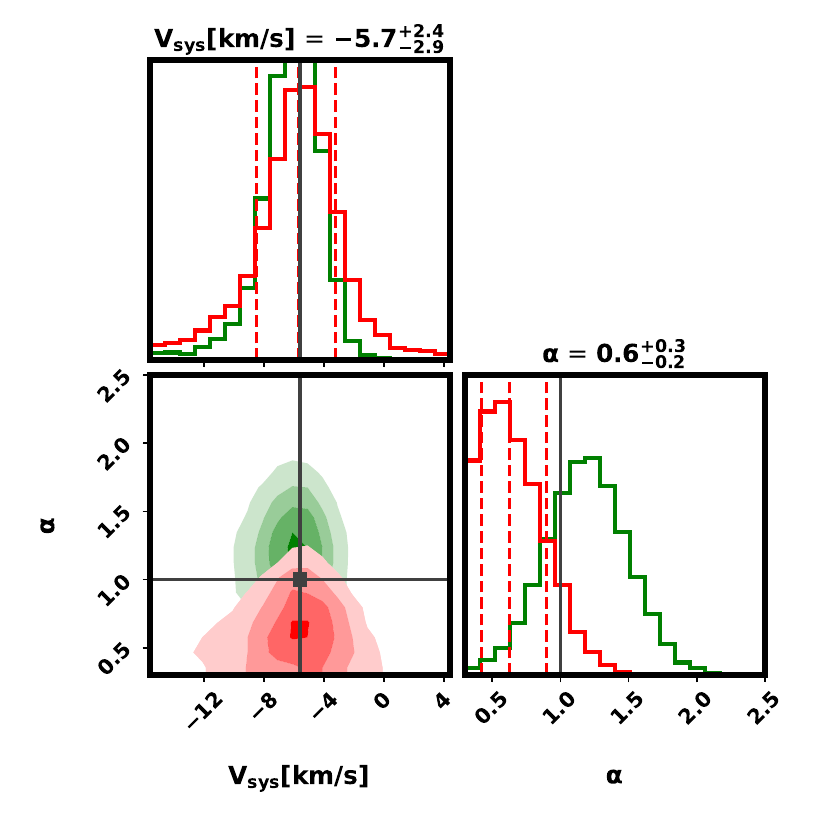}
    \includegraphics[width=0.33\textwidth]{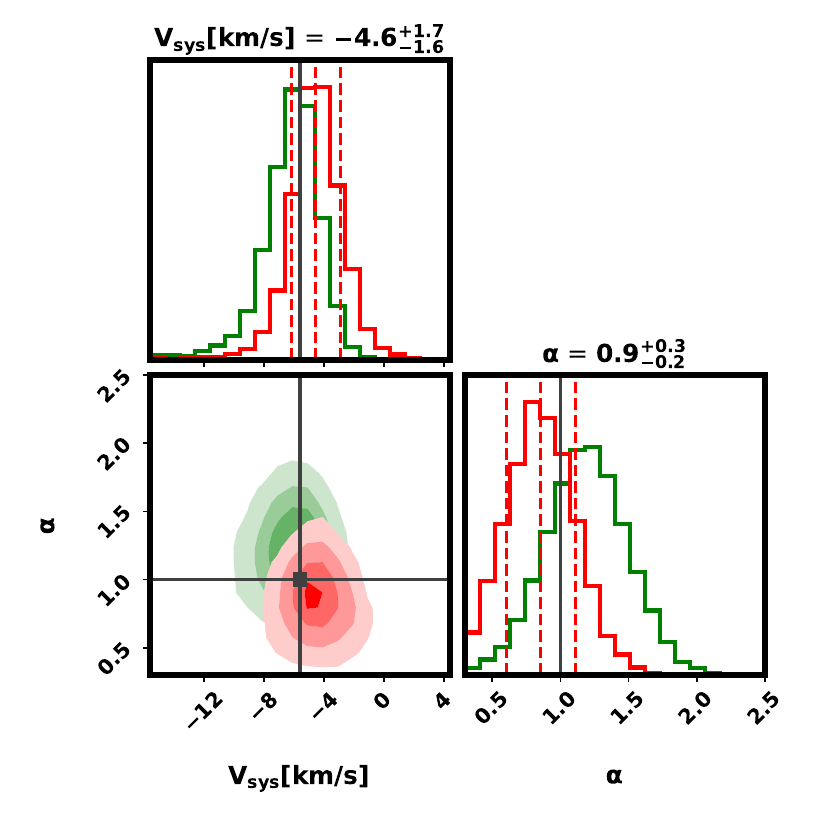}
    \includegraphics[width=0.33\textwidth]{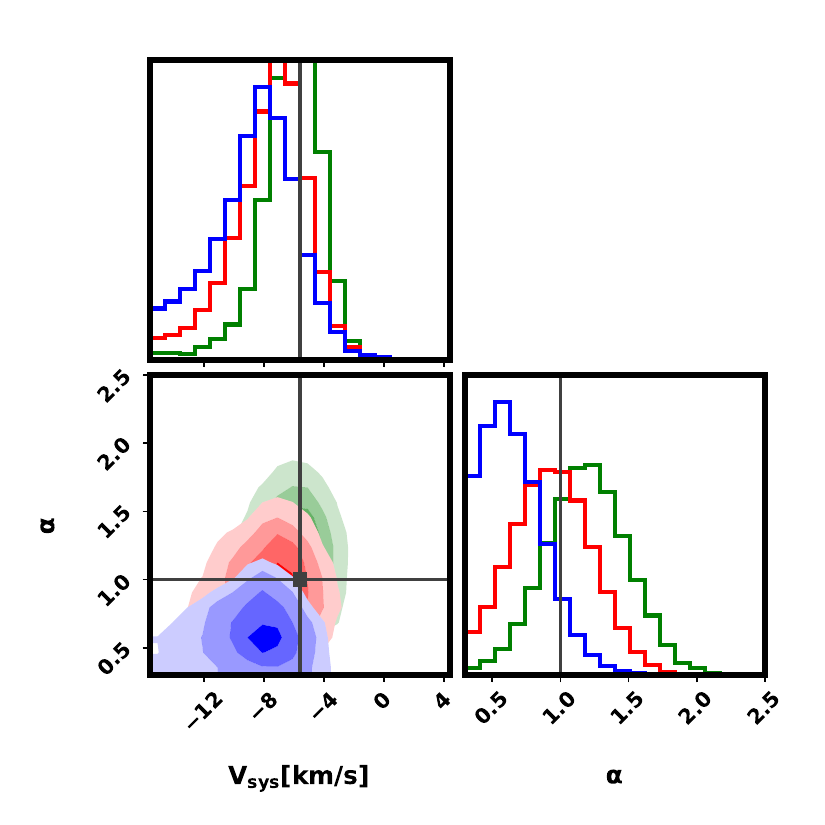}
    \caption{Posterior probability distributions for the inferred tentative detection of \ce{H2S}, under various changes to the methodology. The original posterior probability distribution from Figure \ref{fig:test} is shown in green in all plots, with alternative posterior distributions shown over-plotted (alternative constraints for $V_{\mathrm{sys}}$ and $\alpha$ are printed in some cases). Left: the posterior distribution found when analysing the target spectra which have not been telluric-corrected during reduction with the IGRINS \texttt{PLP}. Middle: the resulting posterior distribution when the time-series model spectra are weighted equally across all in-transit spectra during model comparison, rather than using the weightings generated by \texttt{batman-package}. Right: the calculated posterior distributions when additional principal components are subtracted during detrending - 2 (red) and 3 (blue) rather than the initial 1.
    }
    \label{fig:performance-posteriors}
\end{figure*}

In the preceding work, we analyze spectra which have undergone initial telluric correction during reduction using the IGRINS \texttt{PLP} (Section \ref{clean_norm}). We now instead analyze the target spectra which have not received this correction. We treat the spectra as before, and once again detrend according to the $\Delta$CCF framework, subtracting 1 iteration for \ce{H2S}. A cross-correlation S/N of 1.7 is achieved as before, and a new posterior probability distribution for \ce{H2S} is shown in the left-hand panel of Figure \ref{fig:performance-posteriors} (over-plotted onto what was found previously). 
Although the measured value for $\alpha$ is reduced compared to before, the posterior distribution continues to be constrained with parameters consistent 
with those found previously and the expected parameters for this target e.g. a closely-matching systemic velocity. 
We additionally confirm that we continue to recover a well-constrained posterior for \ce{H2S} when masking fewer spectral orders a priori e.g. when masking the same 12 IGRINS orders as \cite{brogi2023}.

Second, we test the lightcurve correction that is applied to weight the time-series model spectra during model comparison (Figure \ref{fig:observing-conditions}) in the calculation of the posterior probability distribution (Figure \ref{fig:test}). We now repeat the analysis instead using an equal model weighting for each in-transit exposure (i.e. when generating the time-series model template, we inject according to a top-hat profile; see the blue points in the right-hand panel of Figure \ref{fig:observing-conditions}). The new posterior probability distribution is shown in the center panel of Figure \ref{fig:performance-posteriors}, again over-plotted onto what was found previously. The signal for \ce{H2S} persists (with an updated detection significance of 3.0$\sigma$ / $B=22$), although a decreased value for $\alpha$ is recovered without our profile correction ($0.9^{+0.3}_{-0.2}$). This is expected given that the time-integrated model signal is stronger in this case than it should be, and consistent with the $\sim$34\,$\%$ estimated in Section \ref{lightcurve}. Additionally, we test and confirm that the detectabilities of other molecular species determined through injection and recovery tests may be overestimated if the time-series model spectra are not weighted in time by the lightcurve of the planet, for planets with similar grazing transits.

Third, we investigate the stability of our \ce{H2S} inference against the number of principal components subtracted during detrending, to ensure that it is not the spurious result of over-optimized detrending parameters. The posterior probability distributions calculated in different cases are shown in Figure \ref{fig:performance-posteriors}. The posteriors when either 1 (the optimal number from $\Delta$CCF), 2, or 3 principal components are subtracted in detrending are consistent with each other and with the expected parameters. However, we note that observed differences in the calculated $\alpha$ may suggest that the model reprocessing is not as effective as intended, with the median value decreasing when more principal components are subtracted. This may suggest that the erosion of the planet signal in the data is not correctly accounted for here by the model reprocessing. The variation in detection significance is also considered. For example, in the case of subtracting 2 principal components, a detection significance of 2.9$\sigma$ is achieved.

Such testing reveals how small variations can be introduced into the posterior probability distribution as a result of minor changes to the analysis, highlighting the challenge of propagating uncertainties through each and every step in the analysis of high-resolution spectra. As a result, we note that the stated uncertainties on the planetary parameters calculated from the posterior distribution may represent an underestimate of the true uncertainty in the data. Future analysis of similar datasets could aim to systematically quantify the full range of uncertainties arising from each step in the analysis \citep[e.g.][]{savel2024}.

\section{Atmospheric abundance constraints} \label{diversity}

As a result of considering collectively all the analysis and robustness testing performed throughout Section \ref{results}, we tentatively infer the presence of \ce{H2S} in the atmosphere of the super-Earth L 98-59 d with a significance of $\lesssim$3.9$\,\sigma$ ($B=390$).
In this section, 
for a subset of the molecular species for which we have demonstrated detectability or observed tentative signals (\ce{H2S}, \ce{NH3}, and \ce{CH4}), we now explore grids of models in order to place constraints on their atmospheric chemical abundances and the pressure of any cloud deck in the atmosphere of L 98-59 d. Whist the nominal models used in Section \ref{results} assume a cloud-free atmosphere, the atmospheric spectra used in this section are computed in the presence of a cloud deck at varying cloud-top pressures.

\subsection{Constraints on atmospheric \ce{H2S}}

We calculate the detection significance of \ce{H2S} across atmospheric model space, which we parametrize by chemical abundance and cloud-top pressure. We compute a grid of models, for abundances corresponding to 0.1, 1, 10, and 100\,$\times$ solar metallicity ($\mathrm{log_{10}(Z/Z_{\odot})}$ = -1.0, 0.0, 1.0, or 2.0), and for cloud-top pressures of $10^{-2}$, $10^{-3}$, $10^{-5}$ bar, as well as for a cloud-free atmosphere. The abundances explored here are chosen to span a nominal range from depleted to significantly enriched metallicities. For each point in the model grid, we calculate the Bayes factor and detection significance as before, considering the preference for the relevant model over the null model. Each model considered across the grid is compared with the same spectral residuals, detrended according to Table \ref{delta-ccf-table}, allowing for a valid comparison between competing models. However, given that we are now exploring an atmospheric model grid for a single chemical species and aiming to place constraints on the chemical abundance and cloud deck pressure, we are more restrictive on our model prior for $\alpha$ than previously ($0.5<\alpha<1.5$). A tighter prior increases our sensitivity to the absolute spectral line depths which are determined by the abundances/clouds, and therefore allows us to better compare models in the grid.

\begin{figure*}
    \centering
    \includegraphics[width=0.55\textwidth]{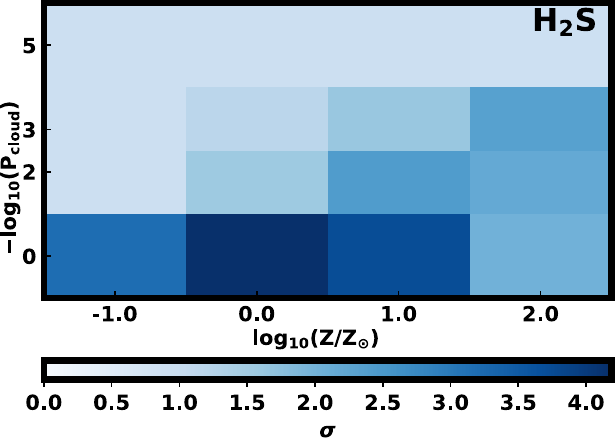}
    \caption{Constraints on the atmospheric abundance of \ce{H2S} and the cloud-top pressure for L 98-59 d. A detection significance is calculated for each atmospheric model in a grid parameterised by metallicity and cloud-top pressure. In terms of the models considered here, we find greatest preference for cloud-free atmospheric models with abundances of \ce{H2S} corresponding to 
    1 and 10\,$\times$ solar metallicity.
    }
    \label{fig:model-constraints}
\end{figure*}

Our results are presented in Figure \ref{fig:model-constraints}. Preference over the null model is demonstrated for atmospheric models of \ce{H2S} for a wide range of chemical abundances and cloud-top pressures. For the models considered, we find that 
cloud-free atmospheric models with abundances of \ce{H2S} corresponding to 1 and 10\,$\times$ solar metallicity in a \ce{H2}-rich atmosphere are most preferred, with significances relative to the null model of 4.2$\sigma$ and 3.7$\sigma$ (Bayes factors of 1240 and 220), respectively. Meanwhile, a cloud-free model with an abundance of \ce{H2S} corresponding to 0.1\,$\times$ solar metallicity is preferred to 3.2$\sigma$ (Bayes factor of 40). This preference for a cloud-free atmosphere is consistent with \cite{banerjee2024}, who also find no evidence for a high-altitude cloud deck.

\subsection{Ruling out other species} \label{section-rule-out}

Using the nominal atmospheric models, we do not find evidence for the presence of \ce{CH4} or \ce{NH3} (Figure \ref{fig:detection-survey_primary}). However, as shown in the bottom panel of Figure \ref{fig:detection-survey_primary}, we are able to recover injected spectra for these molecular species with moderately strong S/N. This demonstrates that, if they were present at their nominal abundances in a cloud-free, \ce{H2}-dominated atmosphere, we would expect to detect them using the current data set. Given this, we now explore the sensitivity of our analysis to the presence of \ce{CH4} and \ce{NH3} considering different chemical abundances and atmospheric cloud-top pressures. In doing so, we calculate the significance to which we can rule out these molecular species in the atmosphere of L 98-59 d, as a function of chemical abundance and cloud-top pressure. To do this, we calculate the preference shown by the data for the null model over the relevant atmospheric model (i.e. calculating the inverse of the previous Bayes factor), for each point in a grid spanning these atmospheric parameters. We here use our initial, wider priors (Table \ref{priors}) in order to provide more conservative estimates for the significances to which we can rule out different atmospheric models. In this way, without using injection/recovery tests, we can quantitatively measure the significance with which we can rule out each atmospheric model given the observed spectra.

\begin{figure*}
    \centering
    \includegraphics[width=0.89\textwidth]{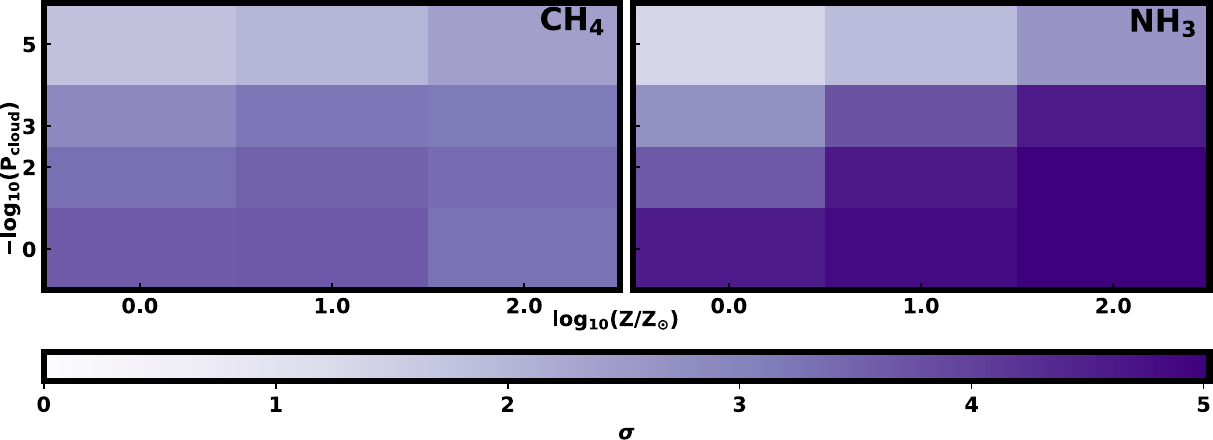}
    \caption{Quantifying the evidence against the presence of \ce{CH4} and \ce{NH3} in the atmosphere of L 98-59 d. Given the sensitivity to \ce{CH4} and \ce{NH3} demonstrated in Figure \ref{fig:detection-survey_primary}, we now calculate the significance to which we can rule out the presence of each of these molecules as a function of chemical abundance and cloud-top pressure.}
    \label{fig:rule-out}
\end{figure*}

For each of \ce{CH4} and \ce{NH3}, this non-detection significance is calculated across an atmospheric model grid given by metallicity and cloud-top pressure (Figure \ref{fig:rule-out}). As shown, for certain chemical abundances and cloud-top pressures, we are able to rule out the presence of these molecules in the atmosphere of L 98-59 d with moderate to high confidence, assuming an \ce{H2}-dominated atmosphere. This may be expected given the recoveries of the model injections for these molecules in Figure \ref{fig:detection-survey_primary}. In the case of a cloud-free atmosphere, favored when placing constraints on \ce{H2S} in Figure \ref{fig:model-constraints}, we are able to rule out \ce{CH4} and \ce{NH3} with significances of $\gtrsim$3.3 -- 3.6$\sigma$ and $\gtrsim$4.6 -- 5$\sigma$, respectively, depending on the relevant chemical abundance. For example, the data prefers the null model, at 3.6$\sigma$ ($B=165$) and 4.8$\sigma$ ($B=20 400$), respectively, over cloud-free atmospheric models containing each of \ce{CH4} and \ce{NH3} with abundances corresponding to 10\,$\times$ solar metallicity. For cloud-free atmospheric models with chemical abundances corresponding to 1\,$\times$ solar metallicity, the respective significances are $3.6\sigma$ ($B=$ 160) and $4.6\sigma$ ($B=$ 6000). Given the preference for a \ce{H2}-dominated, cloud-free atmosphere in order to explain the observed \ce{H2S} (Figure \ref{fig:model-constraints}), our results therefore suggest that the abundances of \ce{CH4} and \ce{NH3} in the atmosphere of L 98-59 d are equivalent to sub-solar metallicities, at confidences of $3.6\sigma$ and $4.6\sigma$, respectively. This is consistent with \cite{gressier2024} and \cite{banerjee2024}, where neither \ce{CH4} and \ce{NH3} are detected in the atmosphere of L 98-59 d.

The data continues to show limited preference for the null model over models of \ce{CH4} and \ce{NH3} even in the presence of a high-altitude cloud deck in the atmosphere of L 98-59 d. For example, this preference is $\sim2.5\sigma$ ($B=5-10$) for models of both molecules with metallicities of 100\,$\times$ solar and a cloud-top pressure of $10^{-5}$. This increases to $3.2\sigma$ ($B=40$) and $4.6\sigma$ ($B=6 500$) for \ce{CH4} and \ce{NH3}, respectively, for a cloud-top pressure of $10^{-3}$. In fact, with this cloud-top pressure of $10^{-3}$, even solar metallicities for \ce{CH4} and \ce{NH3} can be ruled out at $2.7-2.9\sigma$ ($B=12-16$). 
These results somewhat demonstrate the potential for high-resolution spectroscopy to achieve chemical inferences even in the presence of a high-altitude cloud deck, due to the sensitivity of the technique to spectral line cores \citep[e.g.][]{gandhi_seeing_2020}. As a result, ground-based high-resolution transmission spectroscopy offers opportunities to complement JWST observations of similar planets, which alone may only be able to achieve precise abundance constraints for cloud-deck pressures down to $10^{-4}$ bar \citep[e.g.][]{constantinou_characterising_2022}. On the other hand, we find that the sensitivity of our analysis to some atmospheric models in the grid remains very limited. In such cases, the molecules would not be detected even if present, leading to an inability to confidently rule them out. For example, for an atmospheric model of \ce{NH3} with an abundance of 1\,$\times$ solar metallicity and a high-altitude cloud deck at $10^{-5}$ bar, the data shows no meaningful preference for the null model ($B=$ 1.2). In general, the trend across the grid for \ce{NH3} clearly exemplifies the observational degeneracy between the truncation of spectral features due to the presence of a high-altitude cloud deck and significantly depleted atmospheric abundances.

For the other molecular species considered in Section \ref{results}, we are not able to as confidently constrain their abundances, as a result of their reduced detectability demonstrated in Figure \ref{fig:detection-survey_primary}. However, we compute that the data prefers the null model to 2.0\,$\sigma$, 2.3\,$\sigma$, and 2.8\,$\sigma$, respectively, over models of \ce{CO2}, \ce{CO}, and \ce{H2O}, for cloud-free atmospheres and 10\,$\times$ solar metallicity. These values correspond to Bayes factors of 2.5, 4.2, and 12.5 in favour of the null model, respectively. For the atmospheric model with an abundance of \ce{H2O} corresponding to solar metallicity and a cloud-top pressure of $10^{-2}$, the corresponding significance remains 2.0$\sigma$. Again, this is consistent with the results of \cite{gressier2024} and \cite{banerjee2024}, where none of these species are detected in the atmosphere of L 98-59 d.

\section{Summary and discussion} \label{discussion}

In this work, we assess the atmospheric composition of a temperate super-Earth, L 98-59 d, using ground-based high-resolution transmission spectroscopy. We analyze time-series near-infrared spectra obtained using the IGRINS spectrograph on Gemini-S, and demonstrate the sensitivity of ground-based observations of this kind to detecting various molecular species in the atmospheres of super-Earths. L 98-59 d has been the subject of previous atmospheric studies, with tentative inferences of sulfur dioxide \ce{SO2} and hydrogen sulfide \ce{H2S} achieved using JWST \citep{gressier2024, banerjee2024}. We here confirm the presence of \ce{H2S} in the atmosphere of L 98-59 d at $\lesssim$3.9$\,\sigma$ ($B\sim390$), however \ce{SO2} does not have sufficiently strong spectral features in the studied wavelength range to enable a detection. We additionally place constraints on the atmospheric abundances of \ce{H2S} and other common molecular species which are not detected, as well as the cloud-top pressure of any atmospheric cloud deck. Our work suggests preference for a \ce{H2}-dominated, cloud-free atmosphere with an abundance of \ce{H2S} corresponding to $\sim$1--10\,$\times$ solar metallicity, in which case we are also able to constrain abundances of \ce{CH4} and \ce{NH3} equivalent to sub-solar metallicities at $3.6\sigma$ and $4.6\sigma$, respectively.

A central contribution of this work is to demonstrate the feasibility of ground-based high-resolution transmission spectroscopy for the molecular characterization of \ce{H2}-rich atmospheres of temperate super-Earths. We do this through a number of injection and recovery tests (Figure \ref{fig:detection-survey_primary}), a tentative chemical inference of \ce{H2S} in the data (Figures \ref{fig:e-follow-up} and \ref{fig:test}), and by being able to place constraints on the atmospheric abundances of various molecular species as well as the cloud-top pressure (Figures \ref{fig:model-constraints} and \ref{fig:rule-out}). In this way, we show that it is possible using 8\,m-class ground-based facilities to study such atmospheres,  despite small planetary radii and changes in radial velocity during transit. As a result, this work further extends ground-based high-resolution spectroscopy into the regime of lower-mass and more temperate exoplanets.

Our inference of \ce{H2S} in the atmosphere of L 98-59 d supports the previous tentative inference of this molecule made using JWST NIRSpec G395H \citep{gressier2024, banerjee2024}, and is significant for a number of reasons. Detectable quantities of \ce{H2S} without \ce{H2O} is unlikely with equilibrium chemistry \citep{janssen2023, banerjee2024}. Therefore, given the non-detections of \ce{H2O} by \cite{gressier2024}, \cite{banerjee2024}, and in this work (we tentatively rule out the nominal \ce{H2O} model to 2.8$\sigma$), disequilibrium processes in the atmosphere of L 98-59 d may be favoured as the likely source of the tentatively-detected sulfur-bearing species. As a result, our current evidence for \ce{H2S} may further support this suggestion that disequilibrium processes are prominent in the atmosphere and provide a continuous source of \ce{H2S} and potentially other sulfur-bearing species. Possible disequilibrium production mechanisms may include photochemistry \citep{hu2013, tsai2023} or volcanic outgassing from the planet interior \citep{claringbold2023, tsai2024}. However, given the tentative inference of \ce{SO2} and the non-detection of \ce{H2O} by \cite{gressier2024}, \cite{banerjee2024}, and the similar non-detection of \ce{H2O} in this work, the former formation pathway may be unlikely, since \cite{tsai2024} suggest that \ce{H2O} is a necessary reactant in the photochemical production of \ce{SO2}. Therefore, a volcanic outgassing scenario is favored as the source of detectable \ce{H2S} in the atmosphere \citep[e.g.][]{banerjee2024}. Whilst remaining tentative, this indirect inference of volcanic activity in a \ce{H2}-dominated atmosphere allows us to place tighter constraints on the internal structure of L 98-59 d, suggesting a super-Earth planet with a rocky surface, rather than a water world with which volcanic outgassing is inconsistent. 
More generally, our inference for \ce{H2S} - implying the existence of a \ce{H2}-dominated atmosphere around this planet - may be useful for studies of exoplanet formation, evolution, and demographics, given the radius of L 98-59 d ($1.52^{+0.12}_{-0.10}$\,$\mathrm{R_{\oplus}}$) places it toward the edge of the radius valley and at the boundary of those planets which retain their primary \ce{H2}/\ce{He}-dominated atmospheres.

We again note that whilst \cite{banerjee2024} and \cite{gressier2024} additionally find tentative evidence for \ce{SO2}, this molecule has few lines in the wavelength region spanned by IGRINS (instead requiring observation at longer wavelengths) meaning our analysis is insensitive to its presence. 
We also do not find evidence for \ce{H2S} in the spectra taken during a second night of observations. This may be explained by the increased airmass and availability of fewer exposures (fewer than half as many). As a result, injection and recovery tests suggest that the spectra obtained on this night are not conducive to a chemical detection of \ce{H2S} (Figure \ref{fig:detection-survey_secondary}). 
Further high-quality transit observations of this planet, combining transits and employing multi-wavelength/resolution studies, would be beneficial.
For future observations, we note that potential discrepancies between observing visits may arise from atmospheric variability, perhaps driven by the proposed volcanic activity, which could result in a transient presence of molecular species in the upper regions of the planet atmosphere.

Our confirmation of \ce{H2S} reiterates the potential of ground-based high-resolution spectroscopy to complement space-based observations. For example, at the reduced spectral resolution achievable with JWST, \ce{H2S} has broad absorption features which are relatively weak and strongly overlap with several other species, particularly \ce{H2O}. This make it a challenging molecule to detect due to degeneracy with other species and the potential to be masked by the spectral features of \ce{H2O} \citep{janssen2023, tsai2024, banerjee2024}. However, high-resolution spectroscopy is able to break such degeneracies between species by resolving individual spectral lines, allowing for some molecular detections with increased confidence.

Whilst previous ground-based studies have observed smaller planets \citep[e.g.][]{ridden-harper2023}, our work is the first to make a marginal chemical inference in the atmosphere of a sub-Neptune or super-Earth planet using ground-based facilities, and is the first ground-based inference of a sulfur-bearing species in any planet. In addition, the target considered in this work is also the planet with smallest change in radial velocity during transit for which a ground-based chemical inference has been achieved. The atmospheric characterization of super-Earths will become routine in the era of upcoming Extremely Large Telescopes (ELTs) with 25\,m--40\,m mirrors. With up to 25$\times$ larger collecting area, marginal chemical detections of the type presented here will comfortably surpass the 5$\sigma$ threshold to become conclusive detections. Such facilities may also present opportunities for the detection of trace chemical species of smaller abundance, as well as the more detailed three-dimensional characterization of low-mass planets, analogous to what can presently be achieved for hot Jupiter atmospheres with 8\,m-class telescopes. Our current tentative inference of a molecule in the atmosphere of a super-Earth using ground-based facilities may therefore provide an early demonstration for the potential of future ground-based observations to characterize the atmospheres of small, temperate planets. In this regard, this work initiates a new era for the atmospheric characterization of super-Earth planets using high-resolution spectroscopy with ground-based facilities.

\section*{Acknowledgements}

This work is supported by research grants to N.M. from the MERAC Foundation, Switzerland, and the UK Science and Technology Facilities Council (STFC) Center for Doctoral Training (CDT) in Data Intensive Science at the University of Cambridge (STFC grant No. ST/P006787/1). N.M. and C.C. acknowledge support from these sources toward the doctoral studies of C.C. This work is additionally supported by a research grant to C.C. by ANID-BASAL Project FB210003, the Centro de Excelencia en Astrofísica y Tecnologías Afines (CATA), toward the postdoctoral research of C.C. 
We thank the IGRINS team at Gemini-S for conducting the observations and data reduction. N.M. thanks Gregory Mace of the IGRINS team for helpful discussions. C.C. thanks David Ehrenreich for helpful discussions.
Author contributions: N.M. planned the project, and P.R.M., N.M., and S.C. contributed to the successful observing proposal. S.C. and N.M. conducted the atmospheric modeling. C.C. conducted the analysis of data presented in this work and led the writing of the manuscript, with comments from N.M., S.C, and P.R.M.

\section*{Data Availability}

This work is based on IGRINS data obtained through GO Program GS-2021A-Q-216 (PI: P.R. McCullough). Data for this program is publicly available through the Gemini Observatory Archive: \url{https://archive.gemini.edu/searchform/defaults}.

\bibliographystyle{mnras}
\bibliography{references}{}

\appendix

\section{Adjusting the Transit Duration for Cross-Correlation}

\begin{figure}
    \centering
    \includegraphics[width=0.4\textwidth]{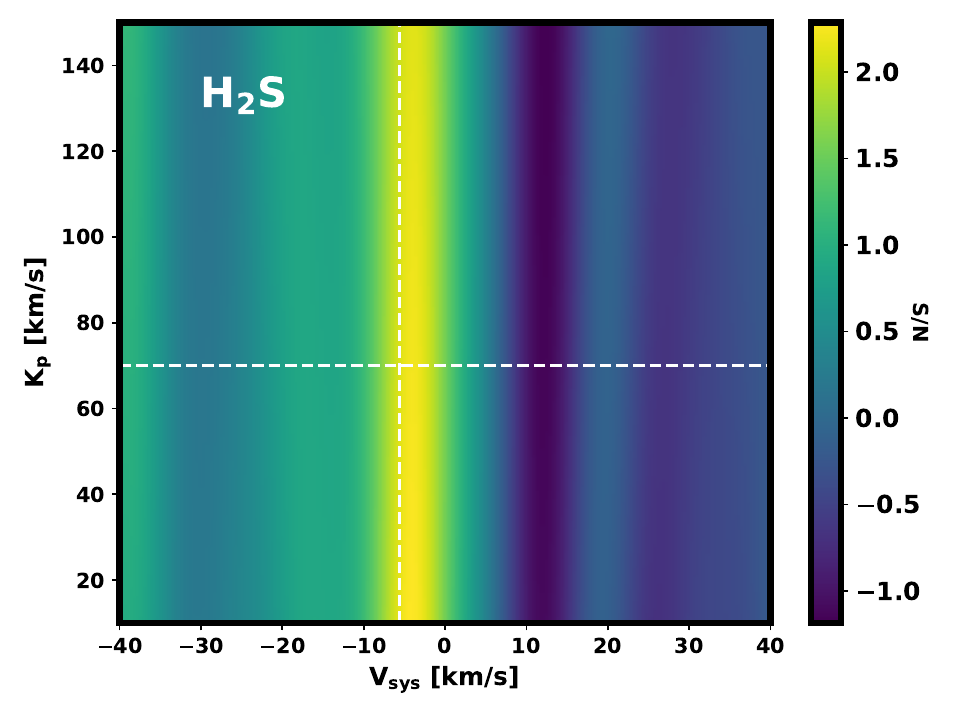}
    \caption{Cross-correlation result for the observed time-series spectra and the model of \ce{H2S}, using an updated value for the transit duration of L 98-59 d. In this case, the transit duration used to calculate the number of in-transit spectra is adjusted according to the left-hand panel of Figure \ref{fig:e-follow-up}, where the optimum number of in-transit spectra is found to be 12 (rather than the 14 calculated using the parameters in Table \ref{system_parameters}). This new value is consistent with the literature value given the associated uncertainties. An updated S/N of 2.3 is calculated.}
    \label{fig:optimum-nin}
\end{figure}

The transit duration for this planet is not well-constrained due to the grazing nature of the transit (Figure \ref{fig:observing-conditions}). The literature value for the transit duration in Table \ref{system_parameters} ($T_{14} = 0.84^{+0.15}_{-0.20}$\,h) is used to calculate the number of in-transit spectra (Table \ref{observing-info}) and subsequently the tentative cross-correlation signal for \ce{H2S} in Figure \ref{fig:detection-survey_primary}. However, Figure \ref{fig:e-follow-up} indicates that a stronger cross-correlation signal is found when fewer exposures (12) are considered to be in-transit. This decreased value is consistent with the uncertainties on the transit duration and is expected given the lightcurve of the planet. Therefore, for completeness, we here show the cross-correlation S/N map calculated when using the smaller number of in-transit spectra (Figure \ref{fig:optimum-nin}), calculating a new S/N for \ce{H2S} of 2.3.

\section{Injection and Recovery Tests}

For planets with small changes in radial velocity during transit, it can be difficult to constrain $K_{\mathrm{p}}$. As a result, for the injection and recovery tests in Figure \ref{fig:detection-survey_primary}, the recovered cross-correlation signals in the S/N maps are degenerate in this parameter. For simplicity and clarity, we here instead show our results for these tests as one-dimensional S/N profiles, taking the cross-section through the expected value for $K_{\mathrm{p}}$ (S/N against systematic velocity). This may aid in assessing the sensitivity of the current dataset to the various molecular species considered in this work. As before, we find that our analysis is moderately sensitive to \ce{CH4} and \ce{NH3}, whilst sensitivity to \ce{H2S}, \ce{H2O}, \ce{CO}, and \ce{CO2} is more tentative.

\begin{figure*}
    \centering
    \includegraphics[width=0.65\textwidth]{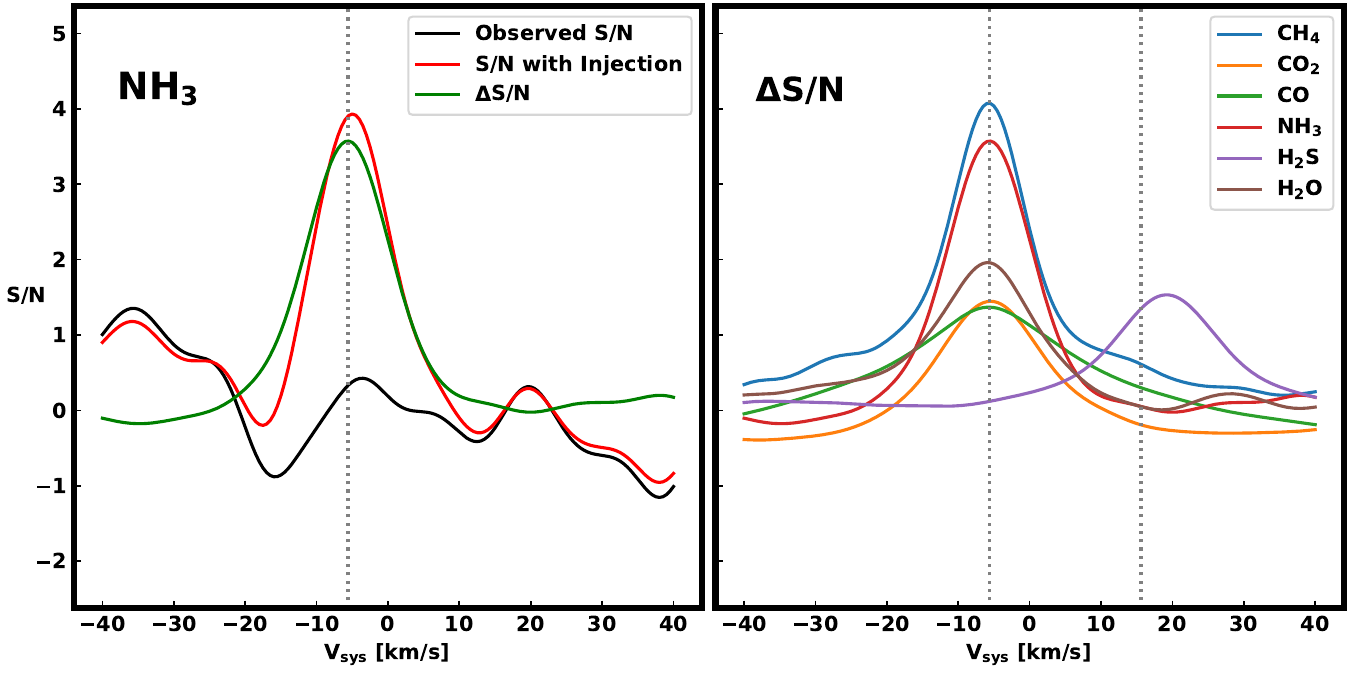}
    \caption{One-dimensional profiles for the cross-correlation S/N recovered during the injection and recovery tests in Figure \ref{fig:detection-survey_primary}. The cross-correlation S/N is taken as the cross-section through the expected value for $K_{\mathrm{p}}$, leaving S/N as a function of $V_{\mathrm{sys}}$. The dotted lines show the velocities of injection. As in Figure \ref{fig:detection-survey_primary}, the injected signal for \ce{H2S} is shifted from the systemic velocity by +19\,$\mathrm{kms^{-1}}$. Left: for \ce{NH3}, the S/N is shown both with and without an injection, with the difference between these two cases ($\mathrm{\Delta S/N}$) also plotted. A clear peak is observed due to the injection. Right: $\mathrm{\Delta S/N}$ is now shown for each of the different models considered in this work. The relative strength of the recovered signal indicates the detectability of each chemical species.}
    \label{fig:1d-profiles}
\end{figure*}

\section{Second night of observation} \label{appendix-night2}

\begin{figure*}
    \centering
    \includegraphics[width=0.8\textwidth]{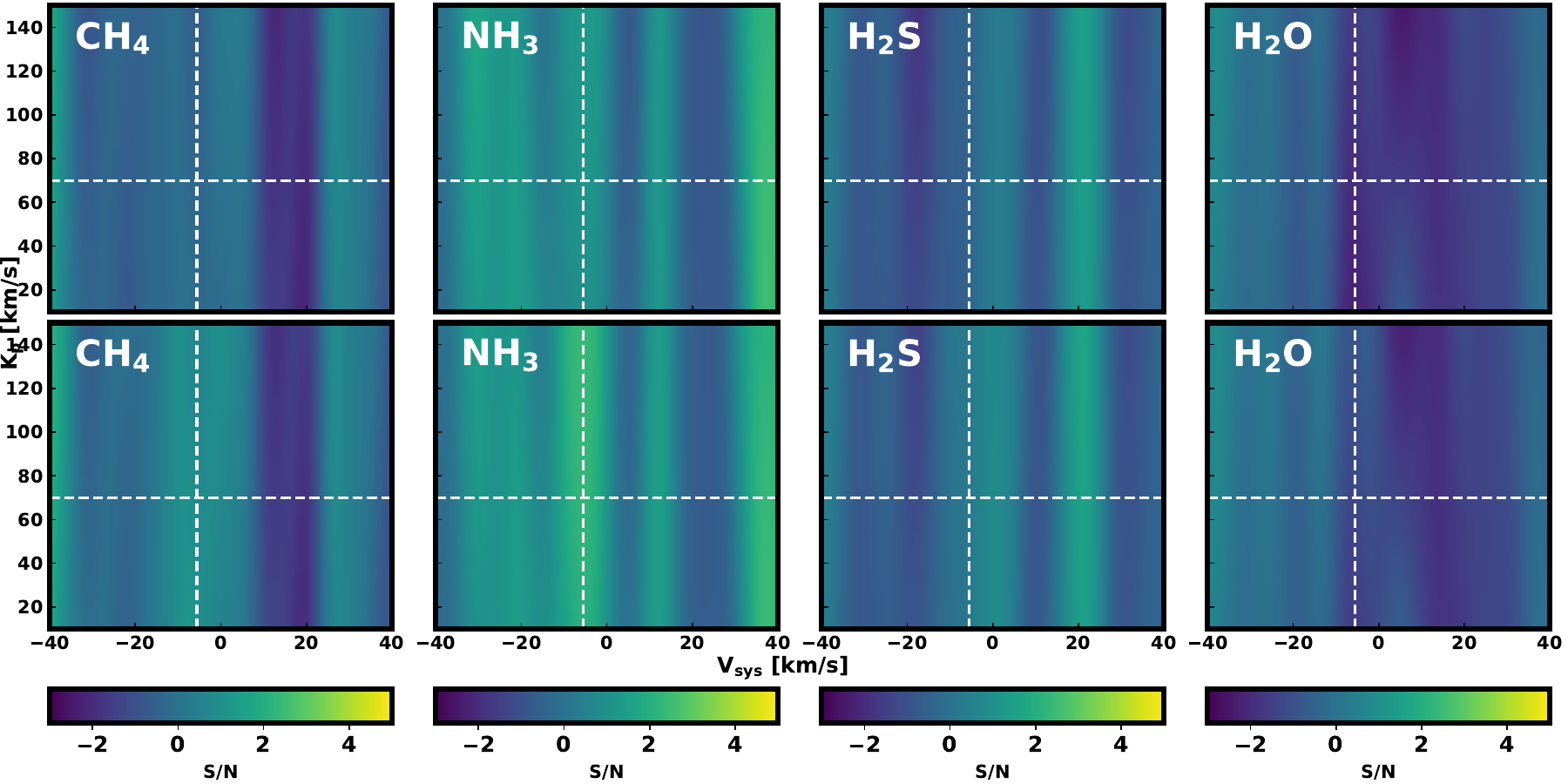}
    \caption{Cross-correlation results for the time-series spectra taken on a second night of observation and a subset of the models considered in this work. As in Figure \ref{fig:detection-survey_primary}, results for the observed spectra are shown in the top panel, whilst injection and recovery tests are conducted in the bottom panel. No significant cross-correlation signals are observed for this night. 
    Injection and recovery testing reveals that this dataset is unlikely to be sensitive to any of the molecular species considered in this work.
    }
    \label{fig:detection-survey_secondary}
\end{figure*}

For the second night of observation (10th February 2021), we do not observe any significant peaks in cross-correlation space, and injection and recovery testing reveals that this dataset is unlikely to be sensitive to any of the molecular species considered in this work (Figure \ref{fig:detection-survey_secondary}). This is perhaps unsurprising given the much reduced number of observations (13 vs 30 exposures) and the increased airmass on this night of observation. We additionally examine this dataset using the likelihood framework (Section \ref{bayesian-methods}) with similar results. We do not combine this night of observation with the primary night, given the demonstrated lack of sensitivity and the potential for planetary atmospheric variability between observing visits likely to undermine efforts in this direction.

\begin{table}
\centering
\begin{tabular}{cc}
\hline \hline
 Model Species & Principal Components \\ \hline
 \hline
 \ce{CH4} & 4 \\ 
 \ce{NH3} & 3 \\ 
 \ce{H2S} & 4 \\ 
 \ce{H2O} & 3 \\ 
 \hline
\end{tabular} 
\caption{The optimum number of principal components to subtract during detrending for spectra observed on the secondary night of observation. These are once more obtained through the $\Delta$CCF framework.}
\label{delta-ccf-table-night2}
\end{table}

\bsp
\label{lastpage}

\end{document}